\documentclass[letterpaper,titlepage,12pt]{article}
\pdfoutput=1
\usepackage{hyperref}

\usepackage{amssymb,amsmath,amsfonts}
\usepackage{epsfig}
\usepackage{mhchem}
\usepackage{tikz}
\usepackage{pgflibraryarrows}
\usepackage{pgflibrarysnakes}

\def\clock{{\count0=\time
          \divide\count0 60
          \ifnum\count0<10 0\fi\the\count0
          \multiply\count0 -60 \advance\count0 \time
          :\ifnum\count0<10 0\fi \the\count0
        }}
\newcommand{\timestamp}{{\small\vbox{\hbox{\tt\jobname.tex}
\hbox{\the\day/\the\month/\the\year, \clock}}}}

\newcommand{\rf}[1]{(\ref{#1})}

\def\be{\begin{equation}}
\def\ee{\end{equation}}
\def\bea{\begin{eqnarray}}
\def\eea{\end{eqnarray}}

\usetikzlibrary{decorations.markings}
\tikzstyle arrowstyle=[scale=1]
\tikzstyle directed=[postaction={decorate,decoration={markings,
    mark=at position .65 with {\arrow[arrowstyle]{stealth}}}}]
\tikzstyle reverse directed=[postaction={decorate,decoration={markings,
    mark=at position .65 with {\arrowreversed[arrowstyle]{stealth};}}}]

\newcommand{\CL}{\mathcal{L}}
\newcommand{\CO}{\mathcal{O}}
\newcommand{\CN}{\mathcal{N}}

\newcommand{\C}{\mathbb{C}}

\newcommand{\ads}{\mbox{AdS}}

\newcommand{\spa}{\ , \ \ }

\newcommand{\ds}{\displaystyle}

\newcommand{\tr}{\mathop{{\rm Tr}}}

\numberwithin{equation}{section}

\begin{document}

\begin{titlepage}

\ \ 

 \vskip 1.5 cm

\centerline{\Huge \bf On Three-point Functions} 
\vskip 0.3cm
\centerline{\Huge \bf in the $AdS_4/CFT_3$ Correspondence}
\vskip 2 cm

\centerline{ {\bf Agnese Bissi$\,^{1,2}$},  {\bf Charlotte Kristjansen$\,^{1}$}, {\bf Ara Martirosyan$\,^{1}$}  and
{\bf Marta Orselli$\,^{1,3}$} }

\vskip 1.1cm

\begin{center}
\sl $^1$ The Niels Bohr Institute, University of Copenhagen \\
\sl  Blegdamsvej 17, DK-2100 Copenhagen \O , Denmark
\vskip 0.3cm
\sl $^2$ The Niels Bohr International Academy, University of Copenhagen  \\
\sl  Blegdamsvej 17, DK-2100 Copenhagen \O , Denmark
\vskip 0.3cm
\sl $^3$ Museo Storico della Fisica e Centro Studi e Ricerche Enrico Fermi,\\
Dipartimento di Fisica and Sezione I.N.F.N., Universit\`a di Perugia,\\
Via Pascoli, I-06123 Perugia, Italy
\end{center}
\vskip 0.7cm

\centerline{\small\tt bissi@nbi.dk, kristjan@nbi.dk, martiros@nbi.dk, orselli@nbi.dk}

\vskip 1.7cm \centerline{\bf Abstract} \vskip 0.2cm \noindent

\noindent
We calculate planar, tree-level, non-extremal three-point functions of 
operators belonging to the $SU(2)\times SU(2)$ sector of ABJM theory.
First, we generalize the determinant representation, found
by Foda for the three-point
functions of the $SU(2)$ sector of \mbox{${\cal N}=4$} SYM,
to the present case and find that, up to normalization factors, the ABJM result 
factorizes into a product of two ${\cal N}=4$ SYM correlation functions.
Secondly, we treat the case where
two operators are heavy and one is light and  BPS, using a coherent
state description of the heavy ones. We show that 
when normalized by the
three-point function of three BPS operators the heavy-heavy-light 
correlation function agrees, in the Frolov-Tseytlin limit,
with its string theory counterpart which we calculate holographically.

\end{titlepage}

\small
\tableofcontents
\normalsize

\setcounter{page}{1}


\section{Introduction}

Unlike what is the case for the $AdS_5\times S^5$-correspondence 
not much is known about  three-point functions of its $AdS_4\times
\C P^3$ cousin. Planar three-point functions of  
scalar chiral primaries were calculated at strong coupling more than 10 years
ago using M-theory on 
$AdS_4 \times S^7$~\cite{Bastianelli:1999en}. More 
recently, strong coupling results were obtained
for the case of
two giant gravitons and one tiny graviton, 
all BPS~\cite{Hirano:2012vz}. These 
three-point functions all show an explicit dependence on the
't Hooft coupling constant and hence are not protected like their 
$AdS_5\times S^5$ counterparts~\cite{Lee:1998bxa,Bissi:2011dc}.~\footnote{In 
\cite{Arnaudov:2010kk} certain three-point functions involving
two (non-BPS)
semi-classical string states and the dilaton field were presented.}
Perhaps for this reason, little effort has 
been put
into studying the corresponding three-point functions at weak coupling.
Weak coupling three-point functions only make sense for operators with
well-defined conformal dimensions i.e.\ for operators which are eigenstates
of the  dilatation operator of the field theory. 
Scalar chiral primary operators belong
to this category. Their two-point functions are 
protected. One can hence immediately proceed with the calculation of 
three-point functions of such operators. A number of tree-level results for
three-point functions of scalar chiral primaries, including
operators dual to giant gravitons, can be found in the 
references~\cite{Hirano:2012vz,Dey:2011ea}. Furthermore, it has been 
shown that the
one-loop correction to any $n$-point function of scalar chiral primaries 
vanishes due to colour combinatorics~\cite{Bianchi:2011rn} 
but apart from that
there are no results on higher loop corrections to correlation
functions neither of chiral primaries nor of more general 
operators.\footnote{It is expected that $n$-point correlation 
functions of BPS operators involving space-time
points with light like separation are related to $n$-sided 
light like polygonal Wilson
loops and to scattering amplitudes~\cite{Bianchi:2011rn} as it is the
case in ${\cal N}=4$ SYM~\cite{Alday:2010zy}.
Similar relations are argued to hold for more general classes of
operators and for theories in general dimensions~\cite{Engelund:2011fg}.}

In the present paper we initiate the study of three-point
functions of scalar operators which are not necessarily chiral primaries.
More precisely, we will be concerned with planar, non-extremal tree-level 
three-point functions of a class of 
operators belonging to the $SU(2)\times SU(2)$
sub-sector. On the field theory side we will exploit the integrability
of the spectral problem~\cite{Minahan:2008hf}
 to represent each operator as a Bethe eigenstate
of an integrable spin chain and then generalize the construction
invented for ${\cal N}=4$ SYM by Escobedo et al.~\cite{Escobedo:2010xs}
and by Foda~\cite{Foda:2011rr}. In addition, we will consider a case
where two of these operators are large and one is small and BPS, and calculate
the corresponding three-point function
in a coherent state approach~\cite{Escobedo:2011xw, Bissi:2011ha}.
The latter three-point function we also determine 
holographically from string
theory using the method developed in~\cite{Zarembo:2010rr}. 
Somewhat surprisingly, if we normalize by 
dividing the result by the 
three-point function of three chiral primaries with matching charges
we obtain the same expression on the string theory
and the gauge theory side.
 
The organization of our paper is as follows. We start by giving a precise
characterization of the operators we wish to consider
in section~\ref{gauge}. Subsequently, in section \ref{foda}, we sketch the derivation
of the three-point 
functions of these operators in the Foda approach~\cite{Foda:2011rr}.
 After that we specialize to the case of two
large and one small BPS operator and determine the three-point function
first from the gauge theory perspective in section~\ref{hhl} and secondly from
the string theory perspective in section~\ref{holo3}. Section \ref{concl} contains our 
conclusion. The details of the Foda approach are given in appendix \ref{appa} 
and  in appendix \ref{appb}  we have collected
the necessary background material on type IIA
strings on $AdS_4\times \C P^3$.

\section{Three-point functions in the $SU(2)\times SU(2)$ sector
of ABJM theory}
\label{gauge}

The field theory which enters the $AdS_4\times \C P^3$ correspondence
\label{abjm}
\cite{Aharony:2008ug} is an $\CN=6$, $U(N)_{k}
\times U(N)_{-k}$ superconformal
Chern-Simons theory. The theory has a 't Hooft expansion with the
't Hooft coupling constant given by $\lambda = N/k$.
Furthermore, it contains two pairs of
chiral superfields transforming in a bi-fundamental representation of
$U(N)\times U(N)$. There is also an $SU(2) \times SU(2)$ R-symmetry
which has been shown to be enhanced to $SU(4)$. 

The scalar sector of the field theory, ABJM theory,
 consists of two complex scalars
$Z_1,Z_2$ which transform in the $N \times \bar{N}$ representation
of $U(N)\times U(N)$ and two complex scalars $W_1,W_2$ which
transform in the $\bar{N}\times N$ representation. The scalars can be
grouped into multiplets of the R-symmetry group $SU(4)$
\begin{equation}
\label{scalars} 
{\cal Z}^a = (Z_1,Z_2,\bar{W}_1,\bar{W}_2) \spa
{\cal \bar{Z}}_a = (\bar{Z}_1,\bar{Z}_2,W_1,W_2),
\end{equation}
with ${\cal Z}^a$ transforming in the fundamental representation and
${\cal \bar{Z}}_a$ in the anti-fundamental representation of $SU(4)$. The
conformal dimension of all the scalars is $\Delta = 1/2$.

A gauge invariant single trace operator containing only scalars is
made by combining the scalars ${\cal Z}^a$ with the scalars ${\cal \bar{Z}}_a$
in an alternating way. Such operators are of the form~\cite{Minahan:2008hf}
\begin{equation}
\label{scalarop} \CO = C^{b_1 b_2 \cdots b_n}_{a_1 a_2 \cdots a_n}
\tr ( {\cal Z}^{a_1} {\cal \bar{Z}}_{b_1} \cdots {\cal Z}^{a_n} 
{\cal \bar{Z}}_{b_n} ).
\end{equation}
The bare dimension of this operator is $n$. Chiral primary operators are
operators for which the tensor $C^{b_1 b_2 \cdots b_n}_{a_1 a_2 \cdots a_n}$
is symmetric in upper as well as lower indices and, in addition, is traceless
when tracing over one upper and one lower index. The spectral problem of
ABJM theory is believed to be 
integrable~\cite{Minahan:2008hf,Gromov:2008qe,Arutyunov:2008if} 
in much the same way as
the spectral problem of ${\cal N}=4$ 
SYM~\cite{Minahan:2002ve,Mandal:2002fs}.
The dilatation operator of the
theory constitutes the Hamiltonian of an integrable spin chain and the
operators with well-defined conformal dimensions are the eigenstates of 
this Hamiltonian. In particular, the scalar operators like~(\ref{scalarop})
have the interpretation of a spin chain state
 of length $2n$ with the spins in the odd
sites transforming in the fundamental and the spins in the even
sites in the anti-fundamental representation of $SU(4)$.

Among the possible sub-sectors of ABJM theory we are interested in the
$SU(2)\times SU(2)$ sector.
This sector is obtained by considering operators made out of 2 scalars 
among ${\cal Z}^a$ and 2 scalars among ${\cal \bar{Z}}_{a}$ 
in Eq.\eqref{scalars} transforming in two separate $SU(2)$
subgroups of $SU(4)$.
If for instance we consider the scalars $Z_{1,2}$ and
$W_{1,2}$, the single-trace operators are of the form
\begin{equation}
\label{su2op} \CO = C^{j_1 j_2 \cdots j_J}_{i_1 i_2 \cdots i_J} \tr
( Z_{i_1} W_{j_1} \cdots Z_{i_J} W_{j_J} ).
\end{equation}
When restricted to the $SU(2)\times SU(2)$ sub-sector the dilatation operator
becomes the Hamiltonian of two decoupled
ferromagnetic $XXX_{1/2}$ Heisenberg spin chains,
one living at the even sites and the other one living at odd sites with the
two chains being  related only through the
momentum constraint~\cite{Minahan:2008hf}. 

In the table below we describe the field content of the three operators
of $SU(2)\times SU(2)$ type which enter the
planar, non-extremal, 
tree-level three-point functions we are interested in.~\footnote{There exist
another class of such three-point functions
which have trivial factorization 
properties~\cite{Pereira:2012}.}

\begin{table}[ht]
\label{genop}
\begin{center}
\begin{tabular}{|c |c | c | c | c |}
\hline Operator  & Vacuum odd & Excitation odd   & 
Vacuum even & Excitation even \\
\hline $\CO_1$ & $(J-J_1)$ ~${Z}_1$ & $J_1$ ~${Z}_2$ & $(J-J_2)$ ~${W}_1$ & 
$J_2$ ~${W}_2$\\
\hline $\CO_2$ &  $(J_1+j_2)$ ~$\bar{Z}_2$ & $(J-J_1-j_1)$ ~$\bar{Z}_1$ & $(J_2+j_2)$ ~$\bar{W_2}$ & 
$(J-J_2-j_1)$ ~$\bar{W}_1$  \\
\hline $\CO_3$ & $j_2$ ~$W_2$ &  
$j_1$ ~$\bar{Z}_1$ &$j_2$ ~$Z_2$  &$j_1$ ~$\bar{W}_1$  \\\hline 
\end{tabular}
\caption{The field content of our
 operators ${\cal O}_1$, ${\cal O}_2$, ${\cal O}_3$
of $SU(2)\times SU(2)$ type 
having a non-vanishing planar, non-extremal three-point function.
The notation $J_1\,\,$ $Z_2$ means that the number of $Z_2$-fields
is $J_1$. It is understood that the number of fields of any type
can not be negative.
}
\end{center}
\end{table}

Here we have indicated which fields are to be considered vacua and which
are to be considered excitations in the interpretation of each operator as 
a state of 
two coupled $XXX_{1/2}$ spin chains. 
We have in mind the situation depicted in
figure~\ref{pant} with site number one being at the left end of each operator.
When we contract the three operators at the planar 
level all vacuum fields from ${\cal O}_3$
are contracted with vacuum fields in ${\cal O}_2$ and all excitations of ${\cal O}_3$ are
contracted with ${\cal O}_1$. This means that only a term in ${\cal O}_3$ 
for which all vacuum fields are to the left of all excitations 
can contribute to the three-point function. Notice also that for contractions
involving ${\cal O}_1$ we connect even sites to even sites and odd sites
to odd sites. For the contractions between ${\cal O}_2$ and ${\cal O}_3$, 
however, odd sites get connected to even sites and vice versa.
We have illustrated the possible contractions in figure~\ref{pant}.
Dashed lines are fields corresponding to excitations and solid lines
are fields corresponding to vacua. The results that we present will
be structure constants $C_{123}$ appearing in the
three-point functions 
\begin{equation}
\langle {O}_1(x) O_2(y)  O_3(z)\rangle=\frac{1}{N}\frac{C_{123}}
{|x-y|^{2(\Delta_1+\Delta_2-\Delta_3)}|x-z|^{2(\Delta_1+\Delta_3-\Delta_2)}
|y-z|^{2(\Delta_2+\Delta_3-\Delta_1)}},
\label{C123}
\end{equation}
of unit normalized operators, i.e.\
operators whose two-point functions fullfill
\begin{equation}
\langle \bar{O}_i(x) O_j(y) \rangle=\frac{\delta_{ij}}{|x-y|^{2\Delta}}.
\label{unitnorm}
\end{equation}
%


\section{The Foda approach}
\label{foda}

\begin{figure}[h]
\begin{center}
\begin{tikzpicture}[>=stealth] 
\draw[] (1,0) -- (4,0);
\draw[] (7,0) -- (10,0);
\draw[] (3.5,4) -- (7.5,4);
\draw[red,dashed] (10,0) -- (7.5,4);
\draw[blue,dashed] (9.5,0) -- (7,4);
\draw[red,dashed] (9,0) -- (6.5,4);
\draw[blue,dashed] (8.5,0) -- (6,4);
\draw[blue] (4,0) arc (180:0:1.5);
\draw[red] (3.5,0) arc (180:0:2);
\draw[red] (1,0) -- (3.5,4);
\draw[blue] (1.5,0) -- (4,4);
\draw[red,dashed] (2,0) -- (4.5,4);
\draw[blue, dashed] (2.5,0) -- (5,4);
\node[] at (9.5,-0.5) {$O_{3_R}$};
\node[] at (7.7,-0.5) {$O_{3_L}$};
\node[] at (1.5,-0.5) {$O_{2_L}$};
\node[] at (3.5,-0.5) {$O_{2_R}$};
\node[] at (5.5,4.5) {$O_{1}$};
\end{tikzpicture}
\end{center}
\caption{The possible contractions between ${\cal O}_1$, ${\cal O}_2$ and
${\cal O}_3$. The full lines represent vacua and the dashed lines represent
excitations. The two different colours illustrate fields in the two
different spin chains.
\label{pant}}
\end{figure}
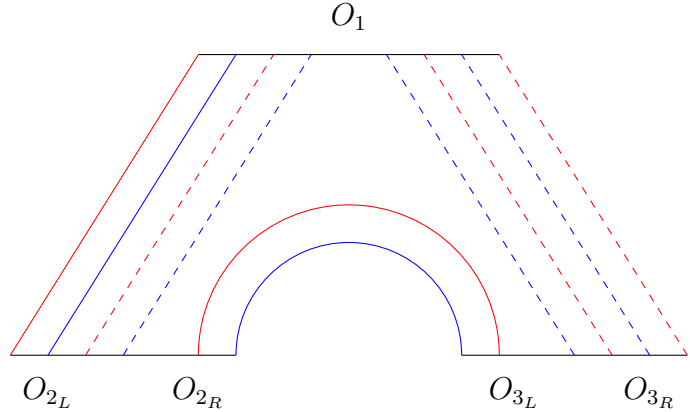

An elegant representation of three-point functions of the $SU(2)$-sector of
${\cal N}=4$ SYM was found by Foda~\cite{Foda:2011rr}. Here, we will generalize
this representation to the $SU(2)\times SU(2)$ sector of ABJM theory. The
key  idea of Foda was to map various parts of
the three-point function onto already known sums over states
for a statistical mechanical lattice model, namely the
6-vertex model. The starting point of Foda's approach is to consider the
operators as spin chain eigenstates as produced by the algebraic
Bethe ansatz. In this picture any given 
eigenstate is obtained from a unit normalized 
reference state (vacuum), which we will take
to be all spins up, by acting with an appropriate series
of spin-flipping or lowering operators. In this picture the 
structure constant corresponding to the three-point
function appearing in figure~\ref{pant} can be written as the following
inner product between Bethe states
\begin{equation}  
C_{123}= {\cal N}_{123}\,
(\,_r\langle {\cal O}_3| \otimes \ce{_{l}}\langle{\cal O}_2| )\, |{\cal O}_1\rangle,
\label{Fodainnerproduct}
\end{equation}
where the subscripts $l$ and $r$ refer to the left and right part respectively
and where ${\cal N}_{123}$ is a normalization constant.
In order to arrive at~(\ref{Fodainnerproduct}) we have exploited the fact
that the inner product between two vacuum states is equal to one. Now 
$|{\cal O}_1\rangle$ is a Bethe eigenstate but
$_r\langle {\cal O}_3| \otimes \ce{_{l}}\langle{\cal O}_2|$ is not. 

In the case of the $SU(2)$-sector of ${\cal N}=4$ SYM the equivalent
of the expression~(\ref{Fodainnerproduct}) could be expressed in terms of
known quantities for the 6-vertex model.
More precisely, the contractions between $|{\cal O}_1
\rangle$
and $|{\cal O}_3\rangle$ gave rise to a factor which
could be identified as a so-called domain wall partition 
function of the 6-vertex model (i.e.\ the partition function of the model
with all initial arrows pointing upwards and all final arrows pointing 
downwards.). What remained was also a quantity which was well-known in the
language of the 6-vertex model, namely another special type of partition
function which could be expressed in terms of a so-called Slavnov
inner product. In the following we will generalize this construction to
the $SU(2)\times SU(2)$ sector of ABJM theory.

 In the ABJM case the operators ${\cal O}_1$, ${\cal O}_2$ and 
${\cal O}_3$ are to be viewed as algebraic Bethe Ansatz
eigenstates of the $SU(2)\times SU(2)$ spin chains and hence must be obtained
from a reference state by acting with a number of spin-flipping operators.
In order to derive these spin-flipping  operators one first
has to construct the necessary $R$-matrices and then form the monodromy matrix.
 The construction
of the four $R$-matrices which are necessary for the full $SU(4)$ spin chain
was carried out 
in~\cite{Minahan:2008hf}. From these $R$-matrices one can form two 
monodromy matrices, one pertaining to
the even sites of the spin chain and the other one to the odd sites of
the spin chain. Consequently, one also gets two sets of lowering operators
$B_e$ and $B_o$ where the subscripts $e$ and $o$ refer to even and odd
respectively. When we constrain to the $SU(2)\times SU(2)$ sub-sector,
two of the four $R$-matrices trivialize and the remaining two become the 
$R$-matrices of two independent $SU(2)$ spin chains, one living on odd
sites and one living on even sites.
Similarly,
the two monodromy matrices simply
become the monodromy matrices of two independent
$SU(2)$ spin chains and finally the
lowering operators $B_e$ and $B_o$ become the usual $SU(2)$ 
spin flipping operators for even and odd sites respectively.
 The two spin-flipping operators $B_o$ and $B_e$ depend on rapidity variables
$\{u_o\}$ and $\{u_e\}$ and in order to obtain an eigenstate both sets of
rapidities $\{u_o\}$ and $\{u_e\}$  have to satisfy the $SU(2)$  
Bethe equations. The only connection
between the two sets of rapidities 
$\{u_o\}$ and $\{u_e\}$ is that they are
related via the momentum constraint which says that the total
momentum of all excitations should vanish and reflects the fact that
the corresponding single trace operator of ABJM theory
should be invariant when one or more pairs of fields are cyclically displaced.
Apart from this constraint we
thus effectively have for each operator
two non-interacting $SU(2)$ spin chains.
In the following we will denote the rapidity variables corresponding to
the operator ${\cal O}_1$ as $(\{u_o\},\{u_e\})$, the rapidity variables 
corresponding to ${\cal O}_2$ as $(\{v_o\},\{v_e\})$, 
and the rapidity variables
corresponding to ${\cal O}_3$ as $(\{w_o\},\{w_e\})$.

Now we can map the elements of each of the two independent $R$-matrices,
the one of the even sites and the one of the odd sites, into the vertex weights
of two independent 6-vertex models. In this way our three-point function
effectively decouples into
 two $SU(2)$ three-point functions.\footnote{The
decoupling is not complete since the cyclicity properties are different for
single trace operators in ${\cal N}=4$ SYM and in ABJM theory.}
 Following
the procedure of Foda~\cite{Foda:2011rr} we can furthermore easily
express the ABJM three-point functions in 
terms of special partition functions of the 6-vertex model.
More precisely, the decoupling properties  imply that we can write
our ABJM three-point function as follows
\begin{eqnarray}
C_{123}=&{\cal N}_{123} &
Z_{j_1}\left( \{w_o\}\right )S[J,J_1,J-J_1-j_1](\{u_o\},\{v_o\})\times 
\nonumber \\
&& Z_{j_1}\left( \{w_e\}\right ) \,
S[J,J_2,J-J_2-j_1](\{u_e\},\{v_e\}).
\label{tp}
\end{eqnarray}
Here the $Z$'s are domain wall partition functions and the $S$'s are Slavnov
inner products. Both types of quantities can be expressed as determinants.
The normalization constant ${\cal N}_{123}^{ABJM}$ takes the form
\begin{equation}
{\cal N}_{123}= \frac{\sqrt{J(j_1+j_2)(J+j_2-j_1)}}{
\sqrt{{\cal N}_{1o} {\cal N}_{1e} {\cal N}_{2o} {\cal N}_{2e}
{\cal N}_{3o} {\cal N}_{3e}}}.
\label{nabjm}
\end{equation}
The quantities in the denominator are the Gaudin norms (i.e.\ the
norms of the eigenstates of the algebraic Bethe ansatz) for the odd and
even parts of the three Bethe states. These norms can also be expressed
as determinants. Finally the factor in the numerator 
takes into account
the cyclic nature of the three operators.\footnote{Notice the difference
to ${\cal N}=4$ SYM that not the full length but half the length of the
operators appear.} 
 In App.~\ref{appa} we will make the arguments of the present section more
precise.~\footnote{After the preparation of this manuscript we learned that
a factorization formula of the same type as~(\ref{tp}) was proposed
(but not substantiated) in~\cite{Pereira:2012}.}


\section{Two heavy and one light operator}

\subsection{The coherent state approach}
\label{hhl}

In this section we wish to calculate a three-point function of the type
considered above in the limit where the two 
operators, ${\cal O}_1$ and ${\cal O}_2$, are much
longer than the operator ${\cal O}_3$.
In order to simplify the presentation we now restrict ourselves to the
following special case\footnote{The general case is not more complicated,
but the notation becomes quite cumbersome.}
\begin{equation}
\label{choice}
j_1=j_2=j~~,~~~J_1=J_2.
\end{equation}
The operator ${\cal O}_3$ then has length $4j$ and
 ${\cal O}_1$ and ${\cal O}_2$ are both 
of the same length, namely $2J$. 
The limit we will be considering is the following
\begin{equation}
1\ll j\ll J_1, J.
\end{equation}
We can represent the long operators
${\cal O}_1$ and ${\cal O}_2$ as coherent states in a $SU(2)\times SU(2)$ spin
chain \cite{Escobedo:2011xw, Bissi:2011ha}. 
The way in which we contract the fields is the same as depicted in 
figure~\ref{pant} but we have to deal with the periodicity of 
the spin chains in a different way. Let us define the first site in 
${\cal O}_1$ for which the corresponding field is contracted with a field in 
${\cal O}_2$ to be site number $2k+2j-1$ of the spin chain corresponding to
${\cal O}_1$. Similarly, let us define the site in the operator ${\cal O}_2$
to which this field is contracted to be site number $2k+2j-1$ of the spin
chain corresponding to ${\cal O}_2$. This in particular means that in
${\cal O}_1$ as well as in ${\cal O}_2$ the fields at the sites
$2k-1,2k,\ldots,2k+2j-2$ are contracted with ${\cal O}_3$.

To take into account all possible contractions
we then have to sum over $k$ from
$k=1$ to $k=J$. We can represent ${\cal O}_1$ in the
following manner
\begin{equation}
{\cal O}_1=\ldots ({\bf u}_o^{(2k-1)}\cdot {\bf Z})
({\bf u}_{e}^{(2k)}\cdot {\bf W})({\bf u}_o^{(2k+1)}\cdot {\bf Z})
({\bf u}_{e}^{(2k+2)}\cdot {\bf W})\ldots
\end{equation}
where the sub-scripts $o$ and $e$ refer to quantities describing the
spin chains at odd and even sites respectively. The vectors
${\bf u}_o=(u_o^1,u_o^2)$ and  ${\bf u}_{e}=(u_{e}^1,u_{e}^2)$ 
belong to $\C^2$ and are unit normalized
, i.e. ${\bf \bar{u}}_o^{(p)}\cdot {\bf u}_o^{(p)}=
{\bf \bar{u}}_{e}^{(p)}\cdot {\bf u}_{e}^{(p)}=1$
and finally ${\bf Z}=(Z_1,Z_2)$, ${\bf W}=(W_1,W_2)$. 
With a similar notation we can write ${\cal O}_2$ as
\begin{equation}
{\cal O}_2=\ldots ({\bf \bar{v}}_o^{(2k-1)}\cdot {\bf \bar{Z}})
({\bf \bar{v}}_{e}^{(2k)}\cdot {\bf \bar{W}})({\bf \bar{v}}_o^{(2k+1)}
\cdot {\bf \bar{Z}})
({\bf \bar{v}}_{e}^{(2k+2)}\cdot {\bf \bar{W}})\ldots
\end{equation}
where $\bar{\bf Z}=(\bar{Z}_1,\bar{Z}_2)$ and
$\bar{\bf W}=(\bar{W}_1,\bar{W}_2)$.
In order for ${\cal O}_1$ and ${\cal O}_2$ to be eigenstates of the
two loop dilatation operator, ${\bf{u}}_o^{(p)}\equiv {\bf{u}}_o(\pi p/J)$
must be periodic in $p$ with period $2J$ and fulfill the
equations of motion of the Landau-Lifshitz sigma model and similarly for
${\bf u}_e$, ${\bf v}_o$ and ${\bf v}_e$~\cite{Grignani:2008is}.\footnote{Notice
that in the present context we are free to choose which fields are considered
vacua and which are considered excitations.}

The third operator, ${\cal O}_3$, is built from
$j$ of each of the fields $Z_1$, $W_1$, $\bar{Z}_2$ and $\bar{W}_2$.
We will now furthermore assume that ${\cal O}_3$ is BPS which implies that
it must be a sum over all
possible orderings of the fields with equal weight. However, only one
ordering of the fields contributes to the planar three-point function, i.e.
\begin{equation}
\CO_{3}  = \CN_3   \tr ((Z_1W_1)^j(\bar{W}_2\bar{Z}_2)^j ) +
\mbox{irrelevant terms},
\end{equation}
where $\CN_3$ is a normalization constant which ensures that the 
two-point function of the operator is unit 
normalized, cf.\ Eq.\eqref{unitnorm}.
 More precisely,
\begin{equation}
\CN_3= \frac{(j!)^2}{\sqrt{(2j)!(2j-1)!}}.
\end{equation}
We can now calculate the planar tree-level three-point function of
our three operators.
The contractions involving ${\cal O}_3$ give rise to the factor 
\begin{equation}
\prod_{m=k}^{k+j-1}
u_o^1\left(\frac{(2m-1)\pi}{J}\right)
u_e^1\left(\frac{2m\pi}{J}\right)
\bar{v}_o^2\left(\frac{(2m-1)\pi}{J}\right)
\bar{v}_e^2\left(\frac{2m\pi}{J}\right),
\end{equation}
and each contraction between ${\cal O}_2$ and ${\cal O}_1$ gives rise
to a factor of ${\bf u}_o\cdot {\bf \bar{v}}_o$ or 
${\bf u}_e\cdot {\bf \bar{v}}_e$. Therefore we can write the three point 
function as\,\footnote{Here we use the notation of~\cite{Escobedo:2010xs}
that each circle in the superscript represents an operator appearing in the
three-point function. Filled circles correspond to non-BPS operators 
and empty circles correspond to
BPS ones.}
\begin{equation}
C^{\bullet \bullet \circ}
={\cal N}_3\,B \sum_{k=1}^J 
\prod_{m=k}^{k+j-1}\frac{u_o^1\left(\frac{(2m-1)\pi}{J}\right)
u_e^1\left(\frac{2m\pi}{J}\right)
\bar{v}_o^2\left(\frac{(2m-1)\pi}{J}\right)
\bar{v}_e^2\left(\frac{2m\pi}{J}\right)
}
{({\bf u}_o^{(2m-1)}\cdot {\bf \bar{v}}_o^{(2m-1)})\,
({\bf u}_e^{(2m)}\cdot {\bf \bar{v}}_e^{(2m)})
},
\end{equation}
where 
\begin{equation}
B= \prod_{m=1}^{J} ({\bf u}_o^{(2m-1)}\cdot {\bf \bar{v}}_o^{(2m-1)})\,
({\bf u}_e^{(2m)}\cdot {\bf \bar{v}}_e^{(2m)}),
\end{equation}
which is the overlap between the operators ${\cal O}_1$ and ${\cal O}_2$.

We now assume that ${\bf u}_{o}$, ${\bf u}_{e}$, 
${\bf v}_{o}$ and ${\bf v}_{e}$ are  
slowly varying, i.e.
${\bf u}_o^{(p)}-{\bf u}_o^{(p-2)}\sim 1/J$ and similarly for 
${\bf u}_{e}$, ${\bf  v}_{o}$ and ${\bf v}_{e}$.  
There is no similar condition relating ${\bf u}_{o}$ and ${\bf u}_{e}$ or
relating ${\bf v}_{o}$ and ${\bf v}_{e}$.
Then we can approximate ${\bf u}_o^{(p)}\equiv{\bf u}_o(\frac{\pi p}{J})$ 
with a 
continuous field ${\bf u}_o(\sigma)$
where $\sigma$ is likewise 
continuous and belongs to the
interval $[0,2\pi]$, and similarly for ${\bf u}_{e}$, ${\bf v}_o$ and
${\bf v}_{e}$.
The statement that ${\cal O}_1$ and ${\cal O}_2$ are
eigenstates of the two-loop dilatation operator now 
translates into the statement
that ${\bf u}_o$, ${\bf u}_{e}$, ${\bf v}_o$ and ${\bf v}_{e}$
obey the continuum Landau-Lifshitz equations of motion.

Due to the fact that the ${\bf u}$'s and ${\bf v}$'s vary slowly  and
in addition that $j\ll J$ we
can now equate all factors in the product over $m$.
Therefore
our three-point function reduces to 
\begin{eqnarray}
C^{\bullet \bullet \circ}&=&
{\cal N}_3\, B \sum_{k=1}^J 
\left(\frac{u_o^1\left(\frac{(2k-1)\pi}{J}\right)
u_e^1\left(\frac{2k\pi}{J}\right)
\bar{v}_o^2\left(\frac{(2k-1)\pi}{J}\right)
\bar{v}_e^2\left(\frac{2k\pi}{J}\right)
}
{({\bf u}_o^{(2k-1)}\cdot {\bf \bar{v}}_o^{(2k-1)})\,
({\bf u}_e^{(2k)}\cdot {\bf \bar{v}}_e^{(2k)})
}\right)^j \cr
&\longrightarrow &
{\cal N}_3\,B J \int_0^{2\pi} \frac{d\sigma}{2\pi} 
\left(\frac{u_o^1(\sigma)
u_e^1(\sigma)
\bar{v}_o^2(\sigma)
\bar{v}_e^2(\sigma)
}
{({\bf u}_o(\sigma)\cdot {\bf \bar{v}}_o(\sigma))\,
({\bf u}_e(\sigma)\cdot {\bf \bar{v}}_e(\sigma))
}\right)^j.
\label{gtleading}
\end{eqnarray}
We now choose ${\cal O}_1$ and ${\cal O}_2$ so similar that 
\begin{equation}
{\bf v}_{(a)}(\sigma)\approx{\bf u}_{(a)}(\sigma)+
\delta {\bf u}_{(a)},
\end{equation}
where $\delta {\bf u}_{(a)}$ is of order $j/J$. A procedure for implementing
this choice at the level of Bethe roots was given in~\cite{Escobedo:2011xw}.
Then, as shown in~\cite{Escobedo:2011xw, Bissi:2011ha}
we get in the limit $j/J\rightarrow 0$ that $B=1$ and our three-point function
can be written as
\begin{equation}
C^{\bullet \bullet \circ}=
{\cal N}_3
\, J \int_0^{2\pi} \frac{d\sigma}{2\pi} 
\left(u_o^1(\sigma)
u_e^1(\sigma)
\bar{u}_o^2(\sigma)
\bar{u}_e^2(\sigma)
\right)^j.
\label{treelevel1}
\end{equation}

We have observed that one obtains an interesting match with string theory
if one considers the following quantity

\begin{equation}
r_{\lambda\ll 1}= \left. \frac{C^{\bullet \bullet \circ}}{C^{\circ \circ \circ}} \right|_{\lambda \ll 1},
\label{rll1}
\end{equation}
where $C^{\circ \circ \circ}$ is the three-point correlation function coefficient for three chiral primaries with the same charges as the operators considered in the numerator.

We can compute three point functions of three chiral primaries by considering
a limit of~\eqref{tp} where all the rapidities go to infinity. 
In~\cite{Kostov:2012jr} it was shown how to perform this limit for operators in $\mathcal{N}$=4 SYM theory and the same strategy can be applied in the
present case.
Adapting the procedure of~\cite{Kostov:2012jr} to our operators in the $SU(2) \times SU(2)$ of ABJM theory
we find
\begin{equation}
\label{chir}
C^{\circ \circ \circ}=
J\sqrt{2 j}\frac{(J-J_1+j)! J_1!((J-j)!)^2j!^2}{(J!)^2(J-J_1)!(J_1-j)!(2j)!}.
\end{equation}
Note that, apart from a different normalization, this is precisely the square of the result of~\cite{Kostov:2012jr} for operators in the $SU(2)$ sector of $\mathcal{N}$=4 SYM theory. 
Taking the limit $J, J_1 \to \infty$ keeping $J-J_1$ large, we have
\begin{equation}
\label{chirprim}
C^{\circ \circ \circ}\sim
\mathcal{N}_3 J s^j,
\end{equation}
where we have defined the quantity $s=\frac{J_1(J-J_1)}{J^2}$.

Using the result \eqref{treelevel1}, we then compute
\begin{equation}
\label{rweak}
r_{\lambda\ll 1}=\frac{1}{s^j}\int_0^{2\pi} \frac{d\sigma}{2\pi} (  u_o^1(\sigma) u_e^1 (\sigma) \bar{u}_o ^2 (\sigma) \bar{u}_e^2  (\sigma))^{j}.
\end{equation}

We will show in the next section that for the ratio $r_{\lambda\gg 1}$ at strong coupling we obtain the same result.

\subsection{The holographic approach}
\label{holo3}

Here we compute the holographic three-point function dual to the 
correlator of two heavy and one light operator considered in Sec.~\ref{hhl} using the prescription of \cite{Zarembo:2010rr}. 
The procedure for this computation has already been outlined in \cite{Escobedo:2011xw, Bissi:2011ha, Georgiou:2011qk} for type IIB string theory on $\ads_5 \times S^5$ and can be easily generalized to type IIA string theory on $\ads_4\times \C P^3$ using the results of Ref.s~\cite{Bastianelli:1999en,  Hirano:2012vz, Bastianelli:1999bm}.

Our convention and notation for the $\ads_4 \times \C P^3$ background for type IIA string theory are explained in appendix~\ref{appb}. Here to parametrize the two two-spheres associated to the two $SU(2)$ sectors contained in $\C P^3$ we use two complex vectors ${\bf{U}}_e(\tau, \sigma)=(U^1_e, U^2_e)$ and ${\bf{U}}_o(\tau, \sigma)=(U^1_o, U^2_o)$.~\footnote{Note that in App.~\ref{appb} we use a different parametrization for the two two-spheres. The two parametrizations are related by a coordinate transformation.}
With this parametrization the results of this section will be directly comparable with the ones of Sec.~\ref{hhl}.

The prescription of \cite{Zarembo:2010rr, Escobedo:2011xw, Bissi:2011ha} gives in our case\,\footnote{Note that the unconventional powers of $\kappa$ are due to a rescaling of the time coordinate (see App.~\ref{appb}).}
\begin{equation}
\label{string2}
{C}^{\bullet \bullet \circ} ={a_j}\lambda^{\frac{3}{4}}\int_{-\infty}^{\infty}d\tau_e\int_{0}^{2\pi}\frac{d\sigma}{2\pi}\frac{Y}{\cosh^{2j} \frac{\tau_e}{\kappa}} 
\left[\frac{3 }{\kappa^2\cosh^2 \frac{\tau_e}{\kappa}}-\frac{1}{\kappa^2}-\frac{\left(\partial_a {\bf \bar U}_e\cdot\partial^a {\bf U }_e+\partial_a {\bf \bar U}_o\cdot\partial^a {\bf U }_o\right)}{2}\right], 
\end{equation}
where we already implemented the gauge choice \eqref{gaugechoice}, we introduced the Euclidean time $\tau_e$ and we defined
\begin{equation}
a_j=\sqrt{\pi(4j+1)}\frac{2^{\frac{1}{4}-2j}(2j+1)!}{j!^2}
~,~~~~Y=\left(U^1_e \bar{U}^2_e U^1_o  \bar{U}^2_o \right)^{j}.
\end{equation}

To compare with the result of Sect.~\ref{hhl} we take the Frolov-Tseytlin limit~\cite{Frolov:2003xy, Kruczenski:2003gt} which in our notation reads~\cite{Bissi:2011ha, Grignani:2008is, Harmark:2008gm} 
\begin{equation}
\label{TMlimit}
 \kappa \rightarrow 0 \spa\frac{1}{\kappa}\partial_{\tau} {\bf U}_{e,o}~~{\rm fixed}\spa \partial_\sigma {\bf U}_{e,o}~~{\rm fixed}.
 \end{equation}

A subclass of solutions that can be mapped to coherent spin chain states at weak coupling is given by considering the  parametrization ${\bf{ U }}_{e,o}(\sigma, \tau)=e^{i\tau/\kappa} {\bf{ u}}_{e,o}(\sigma, \tau)$ with the condition $\bar{{\bf u}}_e\cdot {\bf u}_e=1$ and similarly for ${{\bf u}}_o$. The limit \eqref{TMlimit} becomes
\begin{equation}
\label{TMlimit2}
 \kappa \rightarrow 0 \spa\frac{1}{\kappa}\partial_{\tau} {\bf u}_{e,o}~~{\rm fixed}\spa \partial_\sigma {\bf u}_{e,o}~~{\rm fixed}.
 \end{equation}
The functions 
${\bf{u}}_{e,o}$ are solutions of the Landau Lifshitz equations of motion derived from the action \eqref{LLL} and satisfy the Virasoro condition $\bar{\bf u}_e\cdot \partial_{\sigma}{\bf u}_e+\bar{\bf u}_o\cdot \partial_{\sigma}{\bf u}_o=0$.
Note that in our notation, the energy that one computes using the action \eqref{LLL} goes as $E-J\sim \CO(\lambda/J^2)$. This is due to the rescaling of $t$ in \eqref{tt}. This rescaling has the effect that the gauge constant $\kappa\sim \frac{\sqrt{\lambda}}{J}$. This implies that the expansion in powers of $\kappa$ on the string side parallels the expansion in powers of $\lambda/J^2$ that one has on the gauge theory side.

In the limit \eqref{TMlimit2}, Eq.~\eqref{string2}, to leading order, gives
\begin{equation}
\label{string3}
{C}^{\bullet \bullet \circ} =\lambda^{\frac{3}{4}}\sqrt{\pi(4j+1)}\frac{2^{\frac{1}{4}-2j}(2j+1)!}{j!^2}\int_{-\infty}^{\infty}d\tau_e\int_{0}^{2\pi}\frac{d\sigma}{2\pi} \left(u^1_e \bar{u}^2_e u^1_o  \bar{u}^2_o \right)^{j}\frac{1 }{\kappa^2\cosh^{2+2j} \frac{\tau_e}{\kappa}}.
\end{equation}

For $\kappa \to 0$, the integrand
peaks around $\tau_e=0$ and the $\tau$-integral
can thus be evaluated (see \cite{Escobedo:2011xw, Bissi:2011ha} for more details on this point). The result reads
\begin{equation} 
\label{inttau}
\int_{-\infty}^{+\infty} \frac{d{\tau_e}}{\kappa^2\cosh^{2j+2}  (\frac{\tau_e}{\kappa})} =\frac{2^{2j+1} \left( j! \right)^2}{\kappa\left( 2j+1\right)!}.
\end{equation}

Using that $\kappa = \frac{\sqrt{\lambda}}{J\pi \sqrt{2}}$ (see App.~\ref{appb}) we obtain
\begin{equation}
\label{string4}
{C}^{\bullet \bullet \circ}=J\frac{\lambda^{\frac{1}{4}}2^{\frac{3}{4}}}{ \sqrt{\pi}}\sqrt{4j+1}\int_{0}^{2\pi}\frac{d\sigma}{2\pi}(u^1_e \bar{u}^2_e u^1_o  \bar{u}^2_o)^{j}.
\end{equation}
%


The expression for the holographic three-point function for the chiral primaries with the same charges as the operators considered in Sec.~\ref{hhl} can be computed using Ref.~\cite{Hirano:2012vz}.\,\footnote{Note that, following the notation of Ref.~\cite{Hirano:2012vz}, in our case $p=J-j$. Moreover, from Appendix~A of  \cite{Hirano:2012vz} we have $n_6=j$, $n_1=n_2=p=J-j$, $n_3=j$. Note also that in our notation $\gamma_1=\gamma_2=2j$, $\gamma_3=2J-2j$ and $\gamma=2J+2j$ where we used that the relation between our notation and $J_1$, $J_2$ and $J_3$ in \cite{Hirano:2012vz} is that $(J_1/2)_{\rm there}=(J_2/2)_{\rm there}=J_{\rm our}$ and $(J_3/2)_{\rm there}=2j_{\rm our}$.}
We get
\begin{equation}
\label{string6}
{C}^{\circ \circ \circ}=\frac{\lambda^{\frac{1}{4}}2^{-\frac{1}{4}}}{\sqrt{\pi}}\sqrt{4j+1}\frac{(2J+1)(J-j)!}{(J+j)!}\frac{(J-J_1+j)!}{(J-J_1)!}\frac{J_1!}{(J_1-j)!}.
\end{equation}
Note that this expression differs from \eqref{chir} which is valid at weak coupling. In particular the dependence on the coupling is very different,
showing explicitly
 that the three-point function for three chiral primaries in ABJM theory is not a protected quantity. 
 
In the limit $J, J_1 \to \infty$ with $J-J_1$ large we have
\begin{equation}
\label{string7}
{C}^{\circ \circ \circ}=\frac{\lambda^{\frac{1}{4}}2^{\frac{3}{4}}}{ \sqrt{\pi}}Js^j\sqrt{4j+1}.
\end{equation}

We can now compute the ratio between Eq.~\eqref{string4} and Eq.~\eqref{string7} and compare it with the corresponding quantity \eqref{rweak} at weak coupling. We find
\begin{equation}
r_{\lambda\gg 1}=\left.\frac{C^{\bullet \bullet \circ}}{C^{\circ \circ \circ}}\right|_{\lambda\gg 1}=\frac{1}{s^j}\int_{0}^{2\pi}\frac{d\sigma}{2\pi}(u^1_e \bar{u}^2_e u^1_o  \bar{u}^2_o )^{j}.
\label{rgg1}
\end{equation}

It is easy to see that to leading order we have
\begin{equation}
\label{result}
r_{\lambda\gg 1}=r_{\lambda\ll 1}.
\end{equation}

Note that we have that $r_{\lambda\gg 1}=r_{\lambda\ll 1}$ only in the limit $J, J_1\to \infty$ which is the regime for which also the nice matching of Ref.~\cite{Escobedo:2011xw} was observed.


\section{Conclusion}
\label{concl}

We have seen that the Foda approach to three-point functions generalizes
in a straightforward manner to the $SU(2)\times SU(2)$ sector of ABJM theory.
Obviously a much more challenging project would be to 
extend the approach to the full $SU(4)$ sector.
While the approach of Escobedo et al.\ has been extended to the 
$SO(6)$ sector of $\mathcal N=4$ SYM~\cite{Bissi:2012vx} the Foda approach
has so far resisted generalization, except for the one presented in this
paper and the one of~\cite{Ahn:2012uv} where it was generalized to
spin-1 chains of relevance for certain structure constants in QCD.
 Another interesting line
of investigation would be to include loop corrections. For
ABJM theory three-point functions of chiral
primaries are in general not 
protected so even considering just such operators would 
provide valuable new information. Some progress on
the inclusion of loop corrections in the case of ${\cal N}=4$ SYM
was recently achieved 
in~\cite{Bissi:2011ha, Gromov:2012vu, Grignani:2012yu, Gromov:2012uv, Serban:2012dr}.

In addition, we made the observation that for certain cases
involving two large and
one small and BPS operator one gets agreement between field and string
theory for three-point functions measured relative to three-point functions 
of chiral primaries, to leading order in a large-spin limit. It would be
interesting to investigate if this agreement persists beyond the limit
considered. For this purpose it would be useful to find a way to extract 
the large-spin limit of the heavy-heavy-light correlator from
the Foda approach.\,\footnote{The large-spin 
limit of the heavy-heavy-heavy correlator was extracted from 
the Foda approach in~\cite{Kostov:2012jr}.} Apart from allowing more
directly for a systematic large-spin expansion this would also 
shed light on the connection between
the two different approaches employed in the present work.


\section*{Acknowledgments}

M.O. thanks T. Harmark for useful discussions. A.B., C.K., and A.M. were supported in part by FNU through grant number
272-08-0329.  A.B.\ and C.K.\ acknowledges the kind 
hospitality of Perugia University where parts of this work were done.
In addition, C.K and M.O. would like to thank the organizers of the 
program ``The holographic way: string theory, gauge theory and black holes" 
held at NORDITA, Stockholm where other parts of this work were
carried out.


\begin{appendix}

\section{Details of the Foda approach}
\label{appa}

 As mentioned in the introduction the single trace scalar operators of 
ABJM theory can be viewed as states of a spin chain of even length where the 
variables
on the even sites transform in the fundamental of an $SU(4)$ and the variables
at the odd sites transform in the anti-fundamental of an $SU(4)$. 
The dilatation operator of ABJM theory then acts as a  Hamiltonian for
this spin chain and is conjectured to be integrable.
At the lowest loop order (two loops) this Hamiltonian can be studied by
standard techniques of integrable models~\cite{Minahan:2008hf}. Hence one
can introduce the $R$-matrix, a monodromy matrix and a
transfer matrix. For the alternating $SU(4)$ spin chain one needs a total of
four $R$-matrices~\cite{Minahan:2008hf}
\begin{align}
R_{ab}:V_{a} \otimes V_{b} \longrightarrow V_{a}\otimes V_{b},\hspace{0.5cm}
 R_{ab}(u_o)&=u_o\,I_{a} \otimes I_{b}+\eta P_{ab,} 
\\
R_{\overline{a}\overline{b}} :V_{\overline{a}}\otimes V_{\overline{b}}\longrightarrow V_{\overline{a}}\otimes V_{\overline{b}}\nonumber, \hspace{0.5cm}
R_{\overline{a}\overline{b}}(u_e)&=
u_e\,I_{\overline{a}} \otimes I_{\overline{b}}+\eta P_{\overline{a}\overline{b}}, \\
R_{a\overline{b}} :V_{a}\otimes V_{\overline{b}}\longrightarrow V_{a}\otimes V_{\overline{b}}, \hspace{0.5cm} 
R_{a\overline{b}}(u_o)&=u_o\,I_{{a}} \otimes I_{\overline{b}}+
K_{a\overline{b}},\nonumber \\ 
R_{\overline{a}b} :V_{\overline{a}}\otimes V_{b}\longrightarrow V_{\overline{a}}\otimes V_{b}, \hspace{0.5cm}
R_{\overline{a}b}(u_e)&=u_e\,I_{\overline{a}} \otimes I_{{b}}+K_{\overline{a}b}. \nonumber
\end{align}
Here $V_a$ and $V_{\overline{a}}$ are the vector spaces
of the fundamental and anti-fundamental representation respectively. The 
operator $I$ is the identity operator, $P$ is the permutation, and 
$K$ is the $SU(4)$ trace. Furthermore, $u_e$ and $u_o$ are 
spectral parameters and
$\eta$ is the shift which we will later take to be equal to $i/2$.
From these $R$-matrices one constructs two monodromy matrices, one for sites
of the  fundamental representation and one for sites of the anti-fundamental
representation
\begin{align}
M_{a}(u_{a_{o}})&=R_{a1}(u_{a_{o}})R_{a\overline{1}}(u_{a_{o}})...R_{aJ}(u_{a_{o}})R_{a\overline{J}}(u_{a_{o}}),
\label{monodromy1}\\
M_{\overline{a}}(u_{a_{e}})&=R_{\overline{a}1}(u_{a_{e}})R_{\overline{a}\overline{1}}(u_{a_{e}})...R_{\overline{a}J}(u_{a_{e}})R_{\overline{a}\overline{J}}(u_{a_{e}}).
\label{monodromy2}
\end{align} 
Specializing to the $SU(2)\times SU(2)$ sector the trace operator $K$ does
not contribute and the two $R$-matrices $R_{a\overline{b}}$ and
$R_{\overline{a}b}$ become proportional to the identity. The $R$-matrices
 $R_{ab}$ and
$R_{\overline{a}\overline{b}}$ each become the $R$-matrix of an $SU(2)$
spin chain. We can now generalize the system to an inhomogeneous one
where the $R$-matrices depends on the particular site in question. 
This leads to the following expression for the non-trivial 
$R$-matrices\,\footnote{Here $R_{ab}$ is expressed in the basis 
$(|\uparrow_{a}\rangle \otimes |\uparrow_{b}\rangle, 
|\uparrow_{a}\rangle \otimes |\downarrow_{b}\rangle,
|\downarrow_{a}\rangle \otimes |\uparrow_{b}\rangle,
|\downarrow_{a}\rangle \otimes |\downarrow_{b}\rangle)$ and 
similarly for the other three.}
\begin{align} \label{Rwithoutnorm}
R_{ab}(u_{o},z_{o})&=[u_o-z_o]\,\left(
\begin{array}{cccc}
\frac{[u_{o}-z_{o}+\eta]}{[u_o-z_o]} & 0 & 0 & 0\\
0 & 1 & \frac{[\eta]}{[u_o-z_o]} & 0\\
0 & \frac{[\eta]}{[u_o-z_o]} & 1 & 0\\
0 & 0 & 0 & \frac{[u_{o}-z_{o}+\eta]}{{[u_o-z_o]}}
\end{array}
\right)_{ab}\,\, \equiv [u_o-z_o]\, {\cal R}_{ab},
\\
R_{\overline{a}\overline{b}}(u_{e},z_{e})&=[u_e-z_e]\left(
\begin{array}{cccc}
\frac{[u_{e}-z_{e}+\eta]}{[u_e-z_e]} & 0 & 0 & 0\\
0 & 1 & \frac{[\eta]}{[u_e-z_e]} & 0\\
0 & \frac{[\eta]}{[u_e-z_e]} & 1 & 0\\
0 & 0 & 0 & \frac{[u_{e}-z_{e}+\eta]}{[u_e-z_e]}
\end{array}
\right)_{\overline{a}\overline{b}}\, \,\equiv \, [u_e-z_e]\, 
{\cal R}_{\overline{a}\overline{b}}.
\end{align}
The remaining two are 
\begin{eqnarray}
R_{a\overline{b}}(u_o,z_e)&=&
[u_o-z_e]\,  I,\\ 
R_{\overline{a}b}(u_e,z_o)&=&[u_e-z_o]\,  I.
\end{eqnarray}
Here the parameters $z_e$ and $z_o$ are also denoted as quantum rapidities.
There is one for each site of the spin chain and it is natural to divide
them into two groups, $\{z_o\}$ and $\{z_e\}$, corresponding to respectively
the odd and the even sites. As shown in~\cite{Foda:2011rr} it is convenient to 
keep these parameters arbitrary in the course of the derivation and only
take the homogeneous limit where all $z$'s are identical at the end.

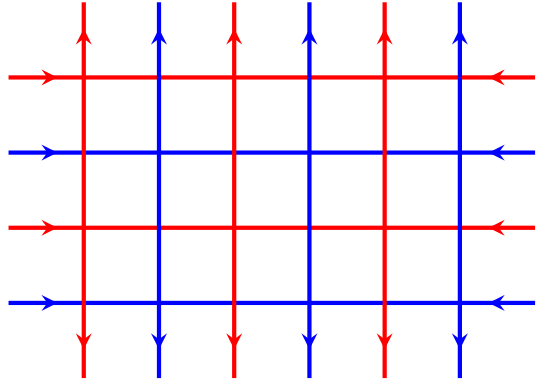
\begin{figure}[h]
\begin{center}
\begin{tikzpicture}[>=stealth] 
\draw[red,ultra thick] (1,4) -- (6,4);
\draw[red,ultra thick, reverse directed] (6,4) -- (7,4);
\draw[blue,ultra thick, directed] (0,3) -- (1,3);
\draw[red,ultra thick,directed] (0,4) -- (1,4);
\draw[blue,ultra thick] (1,3) -- (6,3);
\draw[blue,ultra thick,reverse directed]  (6,3) -- (7,3);
\draw[red,ultra thick, directed] (0,2) -- (1,2);
\draw[red,ultra thick] (1,2) -- (6,2);
\draw[red,ultra thick,reverse directed] (6,2) -- (7,2);
\draw[blue,ultra thick, directed] (0,1) -- (1,1);
\draw[blue,ultra thick] (1,1) -- (6,1);
\draw[blue,ultra thick, reverse directed]  (6,1) -- (7,1);
\draw[red,ultra thick] (1,4) -- (1,1);
\draw[red,ultra thick, directed] (1,4) -- (1,5);
\draw[blue,ultra thick,  directed] (2,4) -- (2,5);
\draw[red,ultra thick, directed] (3,4) -- (3,5);
\draw[blue,ultra thick,  directed] (4,4) -- (4,5);
\draw[red,ultra thick,  directed] (5,4) -- (5,5);
\draw[blue,ultra thick,  directed] (6,4) -- (6,5);

\draw[red,ultra thick, reverse directed] (1,0) -- (1,1);
\draw[blue,ultra thick, reverse directed] (2,0) -- (2,1);
\draw[red,ultra thick, reverse directed] (3,0) -- (3,1);
\draw[blue,ultra thick, reverse directed] (4,0) -- (4,1);
\draw[red,ultra thick, reverse directed] (5,0) -- (5,1);
\draw[blue,ultra thick, reverse directed] (6,0) -- (6,1);

\draw[red,ultra thick] (5,4) -- (5,1);
\draw[red,ultra thick] (3,4) -- (3,1);
\draw[blue,ultra thick] (2,4) -- (2,1);
\draw[blue,ultra thick] (4,4) -- (4,1);
\draw[blue,ultra thick] (6,4) -- (6,1);
\end{tikzpicture}
\end{center}
\caption {A domain wall partition function.
\label{domainwall}}
\end{figure}

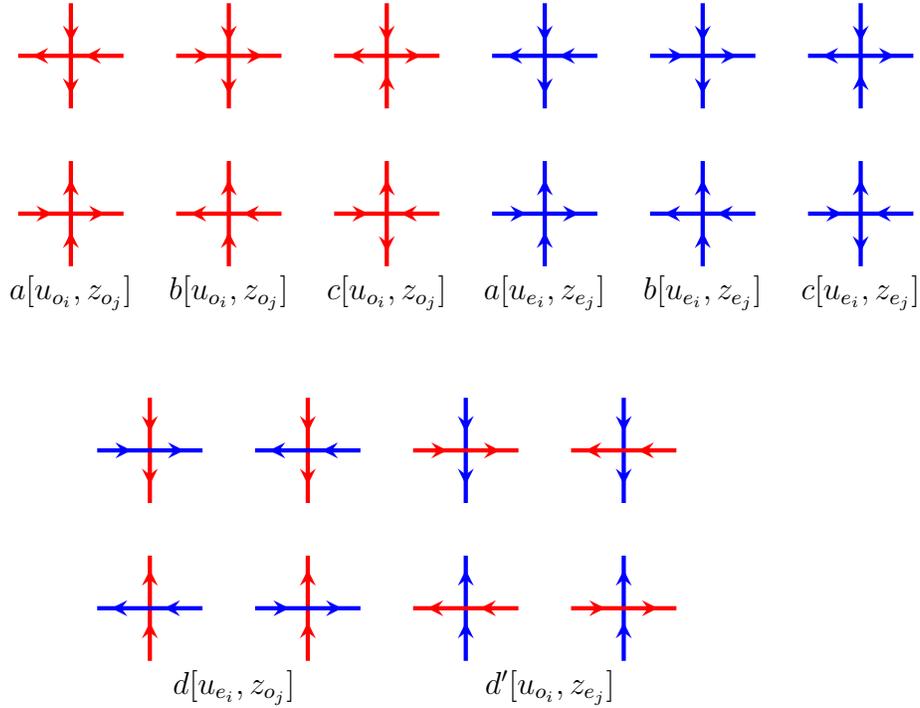
\begin{figure}[htbp]
\begin{center}
\begin{tikzpicture}[scale=0.7]
 \draw[red,ultra thick,directed] (2,0) -- (2,1);
\draw[red,ultra thick,directed] (2,1) -- (2,2);
\draw[red,ultra thick,directed] (1,1) -- (2,1);
\draw[red,ultra thick,directed] (2,1) -- (3,1);

 \draw[red,ultra thick,directed] (5,0) -- (5,1);
\draw[red,ultra thick,directed] (5,1) -- (5,2);
\draw[red,ultra thick,reverse directed] (4,1) -- (5,1);
\draw[red,ultra thick,reverse directed] (5,1) -- (6,1);

\draw[red,ultra thick,reverse directed] (8,0) -- (8,1);
\draw[red,ultra thick,directed] (8,1) -- (8,2);
\draw[red,ultra thick, directed] (7,1) -- (8,1);
\draw[red,ultra thick,reverse directed] (8,1) -- (9,1);

 \draw[blue,ultra thick,directed] (11,0) -- (11,1);
\draw[blue,ultra thick,directed] (11,1) -- (11,2);
\draw[blue,ultra thick,directed] (10,1) -- (11,1);
\draw[blue,ultra thick,directed] (11,1) -- (12,1);

\draw[blue,ultra thick,directed] (14,0) -- (14,1);
\draw[blue,ultra thick,directed] (14,1) -- (14,2);
\draw[blue,ultra thick,reverse directed] (13,1) -- (14,1);
\draw[blue,ultra thick,reverse directed] (14,1) -- (15,1);

\draw[blue,ultra thick,reverse directed] (17,0) -- (17,1);
\draw[blue,ultra thick,directed] (17,1) -- (17,2);
\draw[blue,ultra thick, directed] (16,1) -- (17,1);
\draw[blue,ultra thick,reverse directed] (17,1) -- (18,1);

\draw[red,ultra thick,reverse directed] (2,3) -- (2,4);
\draw[red,ultra thick,reverse directed] (2,4) -- (2,5);
\draw[red,ultra thick,reverse directed] (1,4) -- (2,4);
\draw[red,ultra thick,reverse directed] (2,4) -- (3,4);
\node[] at (2,-0.5) {$a[u_{{o}_{i}},z_{{o}_{j}}]$};
 \draw[red,ultra thick,reverse directed] (5,3) -- (5,4);
\draw[red,ultra thick,reverse directed] (5,4) -- (5,5);
\draw[red,ultra thick,directed] (4,4) -- (5,4);
\draw[red,ultra thick,directed] (5,4) -- (6,4);
\node[] at (5,-0.5) {$b[u_{{o}_{i}},z_{{o}_{j}}]$};
\draw[red,ultra thick,directed] (8,3) -- (8,4);
\draw[red,ultra thick,reverse directed] (8,4) -- (8,5);
\draw[red,ultra thick,reverse directed] (7,4) -- (8,4);
\draw[red,ultra thick,directed] (8,4) -- (9,4);
\node[] at (8,-0.5) {$c[u_{{o}_{i}},z_{{o}_{j}}]$};
 \draw[blue,ultra thick,reverse directed] (11,3) -- (11,4);
\draw[blue,ultra thick,reverse directed] (11,4) -- (11,5);
\draw[blue,ultra thick,reverse directed] (10,4) -- (11,4);
\draw[blue,ultra thick,reverse directed] (11,4) -- (12,4);
\node[] at (11,-0.5) {$a[u_{{e}_{i}},z_{{e}_{j}}]$};
\draw[blue,ultra thick,reverse directed] (14,3) -- (14,4);
\draw[blue,ultra thick,reverse  directed] (14,4) -- (14,5);
\draw[blue,ultra thick,directed] (13,4) -- (14,4);
\draw[blue,ultra thick,directed] (14,4) -- (15,4);
\node[] at (14,-0.5) {$b[u_{{e}_{i}},z_{{e}_{j}}]$};
\draw[blue,ultra thick,directed] (17,3) -- (17,4);
\draw[blue,ultra thick,reverse directed] (17,4) -- (17,5);
\draw[blue,ultra thick,reverse  directed] (16,4) -- (17,4);
\draw[blue,ultra thick,directed] (17,4) -- (18,4);
\node[] at (17,-0.5) {$c[u_{{e}_{i}},z_{{e}_{j}}]$};

\draw[red,ultra thick,directed] (3.5,-2.5) -- (3.5,-3.5);
\draw[red,ultra thick,directed] (3.5,-3.5) -- (3.5,-4.5);
\draw[blue,ultra thick,directed] (2.5,-3.5) -- (3.5,-3.5);
\draw[blue,ultra thick,directed] (3.5,-3.5) -- (4.5,-3.5);

\draw[red,ultra thick,directed] (6.5,-2.5) -- (6.5,-3.5);
\draw[red,ultra thick,directed] (6.5,-3.5) -- (6.5,-4.5);
\draw[blue,ultra thick,reverse directed] (5.5,-3.5) -- (6.5,-3.5);
\draw[blue,ultra thick,reverse directed] (6.5,-3.5) -- (7.5,-3.5);

\draw[blue,ultra thick,directed] (9.5,-2.5) -- (9.5,-3.5);
\draw[blue,ultra thick,directed] (9.5,-3.5) -- (9.5,-4.5);
\draw[red,ultra thick,directed] (8.5,-3.5) -- (9.5,-3.5);
\draw[red,ultra thick,directed] (9.5,-3.5) -- (10.5,-3.5);

\draw[blue,ultra thick,directed] (12.5,-2.5) -- (12.5,-3.5);
\draw[blue,ultra thick,directed] (12.5,-3.5) -- (12.5,-4.5);
\draw[red,ultra thick,reverse directed] (11.5,-3.5) -- (12.5,-3.5);
\draw[red,ultra thick,reverse directed] (12.5,-3.5) -- (13.5,-3.5);

\draw[red,ultra thick,reverse directed] (3.5,-5.5) -- (3.5,-6.5);
\draw[red,ultra thick,reverse directed] (3.5,-6.5) -- (3.5,-7.5);
\draw[blue,ultra thick,reverse directed] (2.5,-6.5) -- (3.5,-6.5);
\draw[blue,ultra thick,reverse directed] (3.5,-6.5) -- (4.5,-6.5);
\node[] at (5.1,-8) {$d[u_{{e}_{i}},z_{{o}_{j}}]$};

\draw[red,ultra thick,reverse directed] (6.5,-5.5) -- (6.5,-6.5);
\draw[red,ultra thick,reverse directed] (6.5,-6.5) -- (6.5,-7.5);
\draw[blue,ultra thick, directed] (5.5,-6.5) -- (6.5,-6.5);
\draw[blue,ultra thick, directed] (6.5,-6.5) -- (7.5,-6.5);

\draw[blue,ultra thick,reverse directed] (9.5,-5.5) -- (9.5,-6.5);
\draw[blue,ultra thick,reverse directed] (9.5,-6.5) -- (9.5,-7.5);
\draw[red,ultra thick,reverse directed] (8.5,-6.5) -- (9.5,-6.5);
\draw[red,ultra thick,reverse directed] (9.5,-6.5) -- (10.5,-6.5);

\draw[blue,ultra thick,reverse directed] (12.5,-5.5) -- (12.5,-6.5);
\draw[blue,ultra thick,reverse directed] (12.5,-6.5) -- (12.5,-7.5);
\draw[red,ultra thick,directed] (11.5,-6.5) -- (12.5,-6.5);
\draw[red,ultra thick,directed] (12.5,-6.5) -- (13.5,-6.5);
\node[] at (11.1,-8) {$d'[u_{{o}_{i}},z_{{e}_{j}}]$};
\end{tikzpicture}
\end{center}
\caption{\label{vertices}
Possible vertices with non-zero weights.
}
\end{figure}

Now, the expressions~\rf{monodromy1} 
and~\rf{monodromy2} for the monodromy matrices turn into
\begin{align}
M_{a}(u_{a_{o}},\{z_{o},z_{e}\}_J)&=\left(
\prod_{i=1}^J[u_{a_o}-z_{i_o}][u_{a_o}-z_{i_e}]\right)
{\cal R}_{a1}(u_{a_{o}},z_{1_o})\ldots {\cal R}_{aJ}(u_{a_{o}},z_{J_o}),\\
M_{\overline{a}}(u_{a_{e}},\{z_{o},z_{e}\}_J)&=
\left(
\prod_{i=1}^J[u_{a_e}-z_{i_o}][u_{a_e}-z_{i_e}]\right)
{\cal R}_{\overline{a}\overline{1}}(u_{a_{e}},z_{1_e})\ldots
{\cal R}_{\overline{a}\overline{J}}(u_{a_{e}},z_{J_e}).
\end{align} 
Notice that (as usual) the indices $a$ and $\overline{a}$ refer to 
auxiliary spaces. We see that up to trivial pre-factors we get one 
monodromy matrix which only involves ${\cal R}$-matrices with fundamental
indices and one monodromy matrix which only involves ${\cal R}$-matrices
with anti-fundamental
indices. Our model has hence decoupled completely into two $SU(2)$
models and we can easily construct the eigenstates of the full 
$SU(2)\times SU(2)$ model
by means of eigenstates of the two $SU(2)$ models. (Of course we have to 
bear in mind that we are only interested
in eigenstates which have cyclic symmetry when 
viewed as $SU(2)\times SU(2)$ states.) 
Let us write 
$M_{a}(u_{a_{o}},\{z_{o},z_{e}\}_J)$ in the following way 
\begin{align}
M_{a}(u_{a_{o}},\{z_{o},z_{e}\}_{J})&=
\left(
\begin{array}{cc}
{A}_{o}(u_{a_{o}},\{z_{o},z_{e}\}_{J}) &  
{B}_{o}(u_{a_{o}},\{z_{o},z_{e}\}_{J})\\
{C}_{o}(u_{a_{o}},\{z_{o},z_{e}\}_{J}) &  
{D}_{o}(u_{a_{o}},\{z_{o},z_{e}\}_{J})
\end{array}\right)_{a} \\
&=\left(
\prod_{i=1}^J[u_{a_o}-z_{i_o}][u_{a_o}-z_{i_e}]\right) 
\left(
\begin{array}{cc}
{\cal A}_{o}(u_{a_{o}},\{z_{o},z_{e}\}_{J}) &  
{\cal B}_{o}(u_{a_{o}},\{z_{o},z_{e}\}_{J})\\
{\cal C}_{o}(u_{a_{o}},\{z_{o},z_{e}\}_{J}) &  
{\cal D}_{o}(u_{a_{o}},\{z_{o},z_{e}\}_{J})
\end{array}\right)_{a}\nonumber 
\end{align}
and similarly for $M_{\overline{a}}(u_{a_{e}},\{z_{o},z_{e}\}_J)$.
Then we define the reference state $|\uparrow_{z_{N}}\rangle$ 
as all spins up, i.e.\
$ |\uparrow_{z_{2J}}\rangle=|\uparrow_{z_{1_{o}}}\rangle\otimes|\uparrow_{z_{1_{e}}}\rangle\otimes...\otimes|\uparrow_{z_{J_{o}}}\rangle\otimes|\uparrow_{z_{J_{e}}}\rangle\ $ and from the usual constructions of the algebraic Bethe ansatz
for the $SU(2)$ spin chain it follows that we can create an eigenstate with
respectively $j_1$ spins at even sites flipped
and $j_2$ spins at odd sites flipped
as follows
\begin{equation}
\prod_{i=1}^{j_{1}} B_{e}(u_{i_{e}},\{z_{o},z_{e}\}_{J})
\prod_{i=1}^{j_2}
B_{o}(u_{i_{o}},\{z_{o},z_{e}\}_{J})|\uparrow_{z_{2J}}\rangle,
\end{equation}
where we have used that $B$ operators pertaining to even and odd sites
commute and where we have to require that $\{u_o\}$ and $\{u_e\}$ 
independently satisfy $SU(2)$ Bethe equations. Now, we are ready to map
our model onto two copies of the 6-vertex model following 
Foda~\cite{Foda:2011rr}.
To illustrate the procedure, let us consider the following transition
amplitude
\begin{equation}
\label{transition}
Z_{2J}(\{u_o,u_e\}_J,\{z_o,z_e\}_J
)=\langle \downarrow_{z_{2J}}|\prod_{i=1}^{J} 
{\cal B}_{e}(u_{i_{e}},\{z_{o},z_{e}\}_{J})
\prod_{i=1}^{J}
{\cal B}_{o}(u_{i_{o}},\{z_{o},z_{e}\}_{J})|\uparrow_{z_{2J}}\rangle.
\end{equation}
This transition amplitude can be understood as a domain wall partition
function for a vertex model as shown
in figure~\ref{domainwall}.
Here a vertical blue line represents an odd spin chain site
and an vertical red line an even spin chain site.
Furthermore, each blue horizontal
line represents a (normalized) 
spin-flipping operator ${\cal B}_o$ and each red horizontal
line represents a normalized spin-flipping operator ${\cal B}_e$. 
We start with all spins pointing up and after application of
$2J$ spin-flipping operators (the horizontal lines) we end with a 
configuration with all spins pointing down. If we ignore
the prefactors in front of the ${\cal R}$'s in 
the monodromy matrices this quantity can be mapped
onto a domain wall partition function of a vertex model with the
vertices shown in figure~\ref{vertices} and the following weights
\begin{align}
a[u_{i},z_{j}]=\frac{u_i-z_j+\eta}{u_i-z_j},\hspace{0.7cm}
c[u_{i},z_{j}]&=\frac{\eta}{u_i-z_j}, 
\end{align}
\begin{align}
b[u_{i},z_{j}]=
d[u_{{e}_{i}},z_{{o}_{j}}]=
d'[u_{{o}_{i}},z_{{e}_{j}}]=1.
\end{align}
In particular, the weights of all the mixed (red-blue) vertices are equal
to one. This means that the partition function of the model factorizes into
a partition function of a red model and a partition function of (an
identical) blue model. Each of these models can be identified as a usual 
6-vertex model.
Summarizing we get for the transition amplitude in~(\ref{transition})
\begin{equation}
Z_{2J}(\{u_o,u_e\}_J,\{z_o,z_e\}_J)=
Z_{J}(\{u_{o}\}_{J},\{z_{o}\}_{J})Z_{J}(\{u_{e}\}_{J},\{z_{e}\}_{J}),
\end{equation}
where $Z_{J}(\{u\}_{J},\{z\}_{J})$ is a domain wall partition function
of the 6-vertex model on a lattice of size $J\times J$
connecting an initial state with all arrows pointing
upwards to a final state with all arrows pointing down.

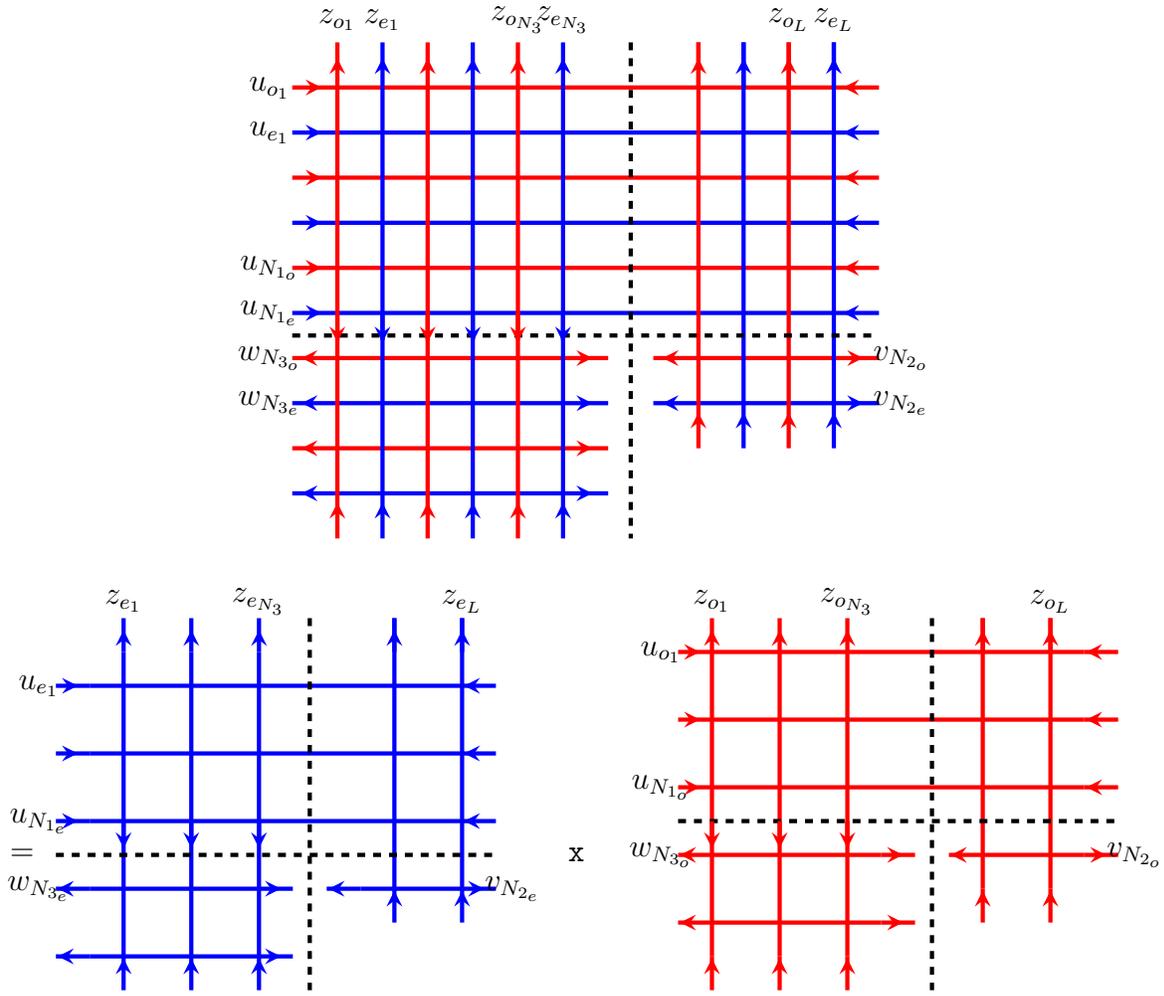
\begin{figure}[htb]
\begin{center}
\begin{tikzpicture}[scale=0.6] 
\draw[red,ultra thick,directed] (0,10) -- (1,10);
\draw[red,ultra thick] (1,10) -- (12,10);
\draw[red, ultra thick,reverse directed] (12,10) -- (13,10);
\draw[blue,ultra thick, directed] (0,9) -- (1,9);
\draw[blue,ultra thick] (1,9) -- (12,9);
\draw[blue, ultra thick,reverse directed] (12,9) -- (13,9);
\draw[red,ultra thick,directed] (0,8) -- (1,8);
\draw[red,ultra thick] (1,8) -- (12,8);
\draw[red, ultra thick,reverse directed] (12,8) -- (13,8);
\draw[blue,ultra thick,directed] (0,7) -- (1,7);
\draw[blue,ultra thick] (1,7) -- (12,7);
\draw[blue, ultra thick,reverse directed] (12,7) -- (13,7);
\draw[red,ultra thick,directed] (0,6) -- (1,6);
\draw[red,ultra thick] (1,6) -- (12,6);
\draw[red, ultra thick,reverse directed] (12,6) -- (13,6);
\draw[blue,ultra thick,directed] (0,5) -- (1,5);
\draw[blue,ultra thick] (1,5) -- (12,5);
\draw[blue,ultra thick, reverse directed] (12,5) -- (13,5);
\draw[red,ultra thick,reverse directed] (0,4) -- (1,4);
\draw[red,ultra thick] (1,4) -- (6,4);
\draw[red,ultra thick, directed] (6,4) -- (7,4);
\draw[blue,ultra thick,reverse directed] (0,3) -- (1,3);
\draw[blue,ultra thick] (1,3) -- (6,3);
\draw[blue,ultra thick,directed]  (6,3) -- (7,3);
\draw[red,ultra thick,reverse directed] (0,2) -- (1,2);
\draw[red,ultra thick] (1,2) -- (6,2);
\draw[red,ultra thick,directed] (6,2) -- (7,2);
\draw[blue,ultra thick,reverse directed] (0,1) -- (1,1);
\draw[blue,ultra thick] (1,1) -- (6,1);
\draw[blue,ultra thick, directed]  (6,1) -- (7,1);
\draw[red,ultra thick,reverse directed] (8,4) -- (9,4);
\draw[red,ultra thick] (9,4) -- (12,4);
\draw[red, ultra thick,directed] (12,4) -- (13,4);
\draw[blue,ultra thick,reverse directed] (8,3) -- (9,3);
\draw[blue,ultra thick] (9,3) -- (12,3);
\draw[blue, ultra thick, directed]  (12,3) -- (13,3);

\draw[red,ultra thick,directed] (1,10) -- (1,11);
\draw[red,ultra thick] (1,10) -- (1,1);
\draw[red, ultra thick,reverse directed] (1,1) -- (1,0);
\draw[blue, ultra thick,directed] (2,10) -- (2,11);
\draw[blue,ultra thick] (2,10) -- (2,1);
\draw[blue, ultra thick,reverse directed]  (2,1) -- (2,0);

\draw[red,ultra thick,directed] (3,10) -- (3,11);
\draw[red,ultra thick] (3,10) -- (3,1);
\draw[red,ultra thick, reverse directed] (3,1) -- (3,0);
\draw[blue, ultra thick,directed] (4,10) -- (4,11);
\draw[blue,ultra thick] (4,10) -- (4,1);
\draw[blue,ultra thick, reverse directed]  (4,1) -- (4,0);

\draw[red,ultra thick,directed] (5,10) -- (5,11);
\draw[red,ultra thick] (5,10) -- (5,1);
\draw[red,ultra thick, reverse directed] (5,1) -- (5,0);
\draw[blue, ultra thick,directed] (6,10) -- (6,11);
\draw[blue,ultra thick] (6,10) -- (6,1);
\draw[blue, ultra thick,reverse directed]  (6,1) -- (6,0);

\draw[red,ultra thick,directed] (9,10) -- (9,11);
\draw[red,ultra thick] (9,11) -- (9,3);
\draw[red, ultra thick,reverse directed] (9,3) -- (9,2);
\draw[blue,ultra thick, directed] (10,10) -- (10,11);
\draw[blue,ultra thick] (10,11) -- (10,3);
\draw[blue, ultra thick,reverse directed]  (10,3) -- (10,2);

\draw[red,ultra thick,directed] (11,10) -- (11,11);
\draw[red,ultra thick] (11,11) -- (11,3);
\draw[red,ultra thick, reverse directed] (11,3) -- (11,2);
\draw[blue,ultra thick,directed] (12,10) -- (12,11);
\draw[blue,ultra thick] (12,11) -- (12,3);
\draw[blue,ultra thick, reverse directed]  (12,3) -- (12,2);
\draw[dashed,ultra thick] (0,4.5)--(13,4.5);
\draw[dashed,ultra thick] (7.5,11)--(7.5,0);
\draw[red,ultra thick,reverse directed] (1,4.4) -- (1,5);
\draw[blue,ultra thick,reverse directed] (2,4.4) -- (2,5);
\draw[red,ultra thick,reverse directed] (3,4.4) -- (3,5);
\draw[blue,ultra thick,reverse directed] (4,4.4) -- (4,5);
\draw[red,ultra thick,reverse directed] (5,4.4) -- (5,5);
\draw[blue,ultra thick,reverse directed] (6,4.4) -- (6,5);
\node[] at (1,11.5) {$z_{o_1}$};
\node[] at (5,11.5) {$z_{o_{N_3}}$};
\node[] at (2,11.5) {$z_{e_1}$};
\node[] at (6,11.5) {$z_{e_{N_3}}$};
\node[] at (12,11.5) {$z_{e_{L}}$};
\node[] at (11,11.5) {$z_{o_{L}}$};
\node[] at (-0.5,10) {$u_{o_{1}}$};
\node[] at (-0.5,9) {$u_{e_{1}}$};
\node[] at (-0.5,6) {$u_{N_{1_o}}$};
\node[] at (-0.5,5) {$u_{N_{1_e}}$};
\node[] at (-0.5,4) {$w_{N_{3_o}}$};
\node[] at (-0.5,3) {$w_{N_{3_e}}$};
\node[] at (13.5,4) {$v_{N_{2_o}}$};
\node[] at (13.5,3) {$v_{N_{2_e}}$};
\end{tikzpicture}
\end{center}
\begin{tikzpicture}[scale=0.45] 
\node[] at (-1,4) {=};
\draw[blue,ultra thick, directed] (0,9) -- (1,9);
\draw[blue,ultra thick] (1,9) -- (12,9);
\draw[blue, ultra thick,reverse directed] (12,9) -- (13,9);
\draw[blue,ultra thick,directed] (0,7) -- (1,7);
\draw[blue,ultra thick] (1,7) -- (12,7);
\draw[blue, ultra thick,reverse directed] (12,7) -- (13,7);
\draw[blue,ultra thick,directed] (0,5) -- (1,5);
\draw[blue,ultra thick] (1,5) -- (12,5);
\draw[blue,ultra thick, reverse directed] (12,5) -- (13,5);
\draw[blue,ultra thick,reverse directed] (0,3) -- (1,3);
\draw[blue,ultra thick] (1,3) -- (6,3);
\draw[blue,ultra thick,directed]  (6,3) -- (7,3);
\draw[blue,ultra thick,reverse directed] (0,1) -- (1,1);
\draw[blue,ultra thick] (1,1) -- (6,1);
\draw[blue,ultra thick, directed]  (6,1) -- (7,1);
\draw[blue,ultra thick,reverse directed] (8,3) -- (9,3);
\draw[blue,ultra thick] (9,3) -- (12,3);
\draw[blue, ultra thick, directed]  (12,3) -- (13,3);

\draw[blue, ultra thick,directed] (2,10) -- (2,11);
\draw[blue,ultra thick] (2,10) -- (2,1);
\draw[blue, ultra thick,reverse directed]  (2,1) -- (2,0);

\draw[blue, ultra thick,directed] (4,10) -- (4,11);
\draw[blue,ultra thick] (4,10) -- (4,1);
\draw[blue,ultra thick, reverse directed]  (4,1) -- (4,0);

\draw[blue, ultra thick,directed] (6,10) -- (6,11);
\draw[blue,ultra thick] (6,10) -- (6,1);
\draw[blue, ultra thick,reverse directed]  (6,1) -- (6,0);

\draw[blue,ultra thick, directed] (10,10) -- (10,11);
\draw[blue,ultra thick] (10,11) -- (10,3);
\draw[blue, ultra thick,reverse directed]  (10,3) -- (10,2);

\draw[blue,ultra thick,directed] (12,10) -- (12,11);
\draw[blue,ultra thick] (12,11) -- (12,3);
\draw[blue,ultra thick, reverse directed]  (12,3) -- (12,2);
\draw[dashed,ultra thick] (0,4)--(13,4);
\draw[dashed,ultra thick] (7.5,11)--(7.5,0);
\draw[blue,ultra thick,reverse directed] (2,4.4) -- (2,5);
\draw[blue,ultra thick,reverse directed] (4,4.4) -- (4,5);
\draw[blue,ultra thick,reverse directed] (6,4.4) -- (6,5);
\node[] at (2,11.5) {$z_{e_1}$};
\node[] at (6,11.5) {$z_{e_{N_3}}$};
\node[] at (12,11.5) {$z_{e_{L}}$};
\node[] at (-0.5,9) {$u_{e_{1}}$};
\node[] at (-0.5,5) {$u_{N_{1_e}}$};
\node[] at (-0.5,3) {$w_{N_{3_e}}$};
\node[] at (13.5,3) {$v_{N_{2_e}}$};
\end{tikzpicture}
\begin{tikzpicture}[scale=0.45] 
\node[] at (-3,4) {\texttt{x}};
\draw[red,ultra thick,directed] (0,10) -- (1,10);
\draw[red,ultra thick] (1,10) -- (12,10);
\draw[red, ultra thick,reverse directed] (12,10) -- (13,10);
\draw[red,ultra thick,directed] (0,8) -- (1,8);
\draw[red,ultra thick] (1,8) -- (12,8);
\draw[red, ultra thick,reverse directed] (12,8) -- (13,8);
\draw[red,ultra thick,directed] (0,6) -- (1,6);
\draw[red,ultra thick] (1,6) -- (12,6);
\draw[red, ultra thick,reverse directed] (12,6) -- (13,6);
\draw[red,ultra thick,reverse directed] (0,4) -- (1,4);
\draw[red,ultra thick] (1,4) -- (6,4);
\draw[red,ultra thick, directed] (6,4) -- (7,4);
\draw[red,ultra thick,reverse directed] (0,2) -- (1,2);
\draw[red,ultra thick] (1,2) -- (6,2);
\draw[red,ultra thick,directed] (6,2) -- (7,2);
\draw[red,ultra thick,reverse directed] (8,4) -- (9,4);
\draw[red,ultra thick] (9,4) -- (12,4);
\draw[red, ultra thick,directed] (12,4) -- (13,4);

\draw[red,ultra thick,directed] (1,10) -- (1,11);
\draw[red,ultra thick] (1,10) -- (1,1);
\draw[red, ultra thick,reverse directed] (1,1) -- (1,0);

\draw[red,ultra thick,directed] (3,10) -- (3,11);
\draw[red,ultra thick] (3,10) -- (3,1);
\draw[red,ultra thick, reverse directed] (3,1) -- (3,0);

\draw[red,ultra thick,directed] (5,10) -- (5,11);
\draw[red,ultra thick] (5,10) -- (5,1);
\draw[red,ultra thick, reverse directed] (5,1) -- (5,0);

\draw[red,ultra thick,directed] (9,10) -- (9,11);
\draw[red,ultra thick] (9,11) -- (9,3);
\draw[red, ultra thick,reverse directed] (9,3) -- (9,2);

\draw[red,ultra thick,directed] (11,10) -- (11,11);
\draw[red,ultra thick] (11,11) -- (11,3);
\draw[red,ultra thick, reverse directed] (11,3) -- (11,2);
\draw[dashed,ultra thick] (0,5)--(13,5);
\draw[dashed,ultra thick] (7.5,11)--(7.5,0);
\draw[red,ultra thick,reverse directed] (1,4.4) -- (1,5);
\draw[red,ultra thick,reverse directed] (3,4.4) -- (3,5);
\draw[red,ultra thick,reverse directed] (5,4.4) -- (5,5);
\node[] at (1,11.5) {$z_{o_1}$};
\node[] at (5,11.5) {$z_{o_{N_3}}$};
\node[] at (11,11.5) {$z_{o_{L}}$};
\node[] at (-0.5,10) {$u_{o_{1}}$};
\node[] at (-0.5,6) {$u_{N_{1_o}}$};
\node[] at (-0.5,4) {$w_{N_{3_o}}$};
\node[] at (13.5,4) {$v_{N_{2_o}}$};
\end{tikzpicture}

\caption{
The decoupling of the three-point function into two parts.\label{threepoint}
}
\end{figure}

Another object of interest for the calculation of three-point functions
is the Slavnov scalar product defined for a single $SU(2)$ spin chain
as
\begin{align} \label{bethescal}
\nonumber
&S[\{u\}_{N_{1}},\{v\}_{N_{2}},\{z\}_{J}]=\\
&=\langle\downarrow_{z_{N_{3},J}}|\prod_{i=1}^{N_{2}} {\cal C}(u_i,\{z\}_{J})
\prod_{j=1}^{N_{1}} {\cal B}(v_{j},\{z\}_{J})|\uparrow_{z_{J}}\rangle,
\end{align}
where 
\begin{equation} \label{instate}
\langle\downarrow_{z_{N_{3},J}}|=
\langle\downarrow_{z_{1}}|\otimes\cdots\otimes\langle\downarrow_{z_{N_{3}}}|
\otimes\langle\uparrow_{z_{N_{3}+1}}|\otimes\cdots\otimes
\langle\uparrow_{z_{J}}|, \nonumber
\end{equation}
with $N_3=N_1-N_2>0$. In the special case where
 $N_1=N_2$, $u_i=v_i$ and $z_i=i/2$ for $i=1,\ldots,N_1$ 
the Slavnov scalar product reduces to
the Gaudin norm, 
\begin{equation}
{\cal N}(\{u\})=S[\{u\}_{N},\{u\}_N,i/2].
\end{equation}
 Generalizing the construction of Foda, a three-point function of the type we 
are interested in can,
up to a normalization factor, be expressed as the partition function of 
the lattice depicted appearing in 
the upper part  of figure~\ref{threepoint}. Again, since the weights
of all vertices of mixed type are equal to one the function factorizes
into a red (even) contribution times a blue (odd) contribution. Each term
is equal to the partition function which one encounters when calculating three
point functions of ${\cal N}=4$ SYM and which was already determined by Foda
who found that it could be written as a product of a Slavnov inner product
and a domain wall partition function both evaluated in the 
homogeneous limit $z_{i_o},z_{i_e}\rightarrow i/2$.
The domain wall partition function comes from the lower left corner
of the lattice while the remaining part constitutes a Slavnov scalar 
product. For simplicity we have depicted a case where we have
the same number of excitations on the odd and the even lattice but
the result holds in the general case as well. Again, it is a simple 
consequence of the decoupling of the two lattices. In order that 
the Bethe eigenstates which enter the three-point functions be normalized
to unity we must divide the result by the Gaudin norm for each operator.
In addition we must multiply by a factor which cures the fact that
the presentation of our three-point function as in figure~\ref{threepoint} 
fails 
to take into account the
cyclicity of the ABJM operators. 
For this final factor 
one does not have a similar complete decoupling into a product of two
factors. This is due  to the alternating nature of 
the ABJM operators which implies that we  only have cyclicity 
(in the horizontal direction) for the 
combined red-blue
model and not for the red and blue model alone. Collecting everything one
gets the expression~(\ref{tp}) for the three-point function.


\section{Type IIA string theory on $AdS_4\times \C P^3$ and
its $SU(2)\times SU(2)$ sigma model limit.}
\label{appb}

The holographic dual of ABJM theory is given by type IIA string theory on $\ads_4\times \C P^3$ \cite{Aharony:2008ug} with metric
\begin{equation}
\label{iiamet}
ds^2 = \frac{R^2}{4} \Big( - \cosh^2 \rho dt^2 + d\rho^2 + \sinh^2
\rho d\hat{\Omega}_2^2 \Big) + R^2 ds_{\C P^3}^2,
\end{equation}
where for the moment we leave the $\C P^3$ part of the metric unspecified and where
\begin{equation}
\label{Rdef} \frac{R^2}{l_s^2} =
\sqrt{2^5 \pi^2 \lambda},
\end{equation}
with $\lambda=N/k$ and with string coupling constant and Ramond-Ramond four-form field strength given by
\begin{equation}
g_s = \Big( \frac{2^5 \pi^2
N}{k^5} \Big)^{\frac{1}{4}}~~,~~~~
F_{(4)} = \frac{3 R^3}{8} \epsilon_{\rm \ads_4}.
\end{equation}
In the regime $\lambda \gg 1$ and $N \ll k^5$, this is a valid background for type IIA string theory \cite{Aharony:2008ug}.

We are interested in zooming in to the $SU(2)\times SU(2)$ sector of type IIA string theory on $\ads_4\times \C P^3$. This can be achieved by taking a limit of small momenta which was first found in \cite{Kruczenski:2003gt} (see also
\cite{Kruczenski:2004kw,Harmark:2008gm, Grignani:2008is, Bissi:2011ha}).
How to do this for type IIA string theory on $\ads_4 \times \C P^3$ is explained in detail in \cite{Grignani:2008is} and the relevant part of the metric becomes
\begin{equation}
\label{theiia} ds^2 = - \frac{R^2}{4} dt^2  + R^2 \Big[ \frac{1}{8}
d\Omega_2^2 + \frac{1}{8} d{\Omega_2'}^2 + (d\delta + \omega)^2
\Big],
\end{equation}
with $R$ given in \eqref{Rdef} and where
\begin{equation}
\begin{array}{c} \ds
\label{defin}
d\Omega_2^2 = d\theta_1^2 + 
\cos^2 \theta_1 d\varphi_1^2\spa d\Omega'_2{}^2 = d\theta_2^2 + 
\cos^2 \theta_2 d\varphi_2^2\\[3mm]
\omega = \frac{1}{4} ( \sin \theta_1 d\varphi_1 +
\sin \theta_2 d\varphi_2)\spa \delta = \frac{1}{4} ( \phi_1+\phi_2-\phi_3-\phi_4 )\\[3mm]
 \varphi_1 =
\phi_1-\phi_2 \spa \varphi_2 = \phi_4-\phi_3
\end{array}
\end{equation}
We see that the coordinates $(\theta_i,\varphi_i)$, $i=1,2$, parametrize two two-spheres corresponding 
to the two $SU(2)$ sectors. For later convenience, the two two-spheres can also be written in terms of two unit vectors fields $\vec{n}_{1,2}$ given by
\begin{equation}
\label{n}
\vec{n}_i=\left(\cos\theta_i \cos\varphi_i, \cos\theta_i \sin\varphi_i, \sin\theta_i \right).
\end{equation}

We now introduce the angular momenta $L_1$ and $L_2$ in one $SU(2)$ and $L_3$ and $L_4$ in the other $SU(2)$ with the condition $L_1+L_2+L_3+L_4 = 0$ .
As explained in \cite{Grignani:2008is} the $SU(2)\times SU(2)$ sector is obtained by  considering
states for which $\Delta - L_1 - L_2$ is small, where $\Delta$ is the energy. 
This can be implemented as a sigma-model limit with the following coordinate transformation
\begin{equation}
\label{tt} \tilde{t} = \lambda' t \spa \chi = \delta -
\frac{1}{2} t,
\end{equation}
where $\lambda'=\lambda/J^2$, $J \equiv L_1+L_2$ 
and so that
\begin{equation}
\label{ham}
\tilde{H} \equiv i \partial_{\tilde{t}} = \frac{( \Delta - J )}{\lambda'} \spa 2J
= - i\partial_\chi,
\end{equation}
We see that sending $\lambda'\to 0$, one has that $\Delta-J \to 0$ which means that we keep the modes of the
$SU(2)\times SU(2)$ sector dynamical, while the other modes become non-dynamical and decouple in this limit.

Using \eqref{tt}, the type IIA metric becomes
\begin{equation}
\label{startmet}
ds^2 = R^2 \left[ ( \frac{1}{\lambda'}d\tilde{t} + d\chi+\omega ) ( d\chi  +
\omega )  +  \frac{1}{8} d\Omega_2^2 + \frac{1}{8} d{\Omega_2'}^2
\right].
\end{equation}
The bosonic sigma-model Lagrangian and Virasoro constraints are
\begin{equation}
\label{lagr} \CL = - \frac{1}{2} G_{\mu \nu} h^{\alpha \beta}
\partial_\alpha x^{\mu} \partial_\beta x^{\nu},
\end{equation}
\begin{equation}
G_{\mu\nu} ( \partial_\alpha x^\mu \partial_\beta x^\nu -
\frac{1}{2} h_{\alpha\beta} h^{\gamma\delta} \partial_\gamma x^\mu
\partial_\delta x^\nu ) = 0,
\end{equation}
with $G_{\mu\nu}$ being the metric \eqref{startmet}.
$h^{\alpha\beta}= \sqrt{-\det \gamma} \gamma^{\alpha\beta}$ with $\gamma_{\alpha\beta}$ being the world-sheet metric. 
 
Our gauge choice is
\begin{equation}
\label{gaugechoice}
\tilde{t} = \kappa \tau,
\end{equation}
\begin{equation}
\label{gaugecon} 2\pi  p_- = \frac{\partial \CL}{\partial
\partial_\tau x^-} = \mbox{const.} \spa \frac{\partial \CL}{\partial
\partial_\sigma x^-} = 0.
\end{equation}
Moreover, the constant $\kappa$ can also be determined from
\begin{equation}
2 J = P_{\chi} = \int_0^{2\pi} d\sigma p_{\chi} = \frac{R^2 \kappa}{2
\lambda'} = \frac{2\pi \sqrt{2\lambda} \kappa}{\lambda'}.
\end{equation}
We see that
$\kappa = \frac{\sqrt{\lambda'}}{\pi \sqrt{2}}$.
Thus $\kappa \to 0$ for $\lambda' \rightarrow 0$. 
Moreover, from \eqref{ham} we have that the right energy scale is given by $\tilde{\tau} = \kappa \tau$. This means that the quantity that we keep fixed in the limit $\kappa \rightarrow
0$ is  $\dot{x}^\mu=\partial_{\tilde{\tau}} x^\mu$.

Proceeding as in \cite{Grignani:2008is}, we can then solve the Virasoro constraints and the gauge conditions order by order in $\kappa$.  This actually corresponds, on the gauge theory side, to an expansion in powers of $\lambda'$. Here we skip the various steps and report the final result for the action to leading order
\begin{equation}
\label{LLL} I = \frac{J}{4\pi} \sum_{i=1}^2 \int d\tilde{t}
\int_0^{2\pi} d\sigma \Big[ \sin \theta_i \dot{\varphi}_i - \pi^2
 ( \vec{n}_i)^2
\Big],
\end{equation}
\begin{equation}
\label{momm} \sum_{i=1}^2 \int_0^{2\pi} d\sigma \sin \theta_i \varphi_i' =0,
\end{equation}
where the last expression gives the momentum constraint.

We see that, up to the perturbative order we are interested in, by taking the $SU(2)\times SU(2)$ sigma-model limit we obtain two Landau-Lifshitz models added together \eqref{LLL}, one for each $SU(2)$, which are related only through the momentum constraint \eqref{momm} \cite{Grignani:2008is}. This is moreover consistent with results on the gauge theory side.

\end{appendix}


\providecommand{\href}[2]{#2}\begingroup\raggedright\endgroup


\begin{thebibliography}{10}

\bibitem{Bastianelli:1999en}
F.~Bastianelli and R.~Zucchini, ``{Three point functions of chiral primary
  operators in d = 3, N=8 and d = 6, N=(2,0) SCFT at large N},''
  \href{http://dx.doi.org/10.1016/S0370-2693(99)01179-X}{{\em Phys.Lett.} {\bf
  B467} (1999)  61--66},
\href{http://arxiv.org/abs/hep-th/9907047}{{\tt arXiv:hep-th/9907047
  [hep-th]}}.

\bibitem{Hirano:2012vz}
S.~Hirano, C.~Kristjansen, and D.~Young, ``{Giant Gravitons on  {$\ads_4 \times \C P^3$}
  and their Holographic Three-point Functions},'' {\em JHEP} {\bf 1207}, 006 (2012),
\href{http://arxiv.org/abs/1205.1959}{{\tt arXiv:1205.1959 [hep-th]}}.

\bibitem{Lee:1998bxa}
S.~Lee, S.~Minwalla, M.~Rangamani, and N.~Seiberg, ``{Three point functions of
  chiral operators in D = 4, N=4 SYM at large N},'' {\em Adv.Theor.Math.Phys.}
  {\bf 2} (1998)  697--718,
\href{http://arxiv.org/abs/hep-th/9806074}{{\tt arXiv:hep-th/9806074
  [hep-th]}}.

\bibitem{Bissi:2011dc}
A.~Bissi, C.~Kristjansen, D.~Young, and K.~Zoubos, ``{Holographic three-point
  functions of giant gravitons},''
  \href{http://dx.doi.org/10.1007/JHEP06(2011)085}{{\em JHEP} {\bf 1106} (2011)
   085}, \href{http://arxiv.org/abs/1103.4079}{{\tt arXiv:1103.4079 [hep-th]}}.
P.~Caputa, R.~d.~M. Koch, and K.~Zoubos, ``{Extremal versus Non-Extremal
  Correlators with Giant Gravitons},''
  \href{http://dx.doi.org/10.1007/JHEP08(2012)143}{{\em JHEP} {\bf 1208} (2012)
   143},
\href{http://arxiv.org/abs/1204.4172}{{\tt arXiv:1204.4172 [hep-th]}}.
H.~Lin, ``{Giant gravitons and correlators},''
\href{http://arxiv.org/abs/1209.6624}{{\tt arXiv:1209.6624 [hep-th]}}.



\bibitem{Arnaudov:2010kk}
D.~Arnaudov and R.~Rashkov, ``{On semiclassical calculation of three-point
  functions in  {$\ads_4 \times \C P^3$}},''
  \href{http://dx.doi.org/10.1103/PhysRevD.83.066011}{{\em Phys.Rev.} {\bf D83}
  (2011)  066011},
\href{http://arxiv.org/abs/1011.4669}{{\tt arXiv:1011.4669 [hep-th]}}.

\bibitem{Dey:2011ea}
T.~K. Dey, ``{Exact Large $R$-charge Correlators in ABJM Theory},''
  \href{http://dx.doi.org/10.1007/JHEP08(2011)066}{{\em JHEP} {\bf 1108} (2011)
   066},
\href{http://arxiv.org/abs/1105.0218}{{\tt arXiv:1105.0218 [hep-th]}}.
S.~Chakrabortty and T.~K. Dey, ``{Correlators of Giant Gravitons from dual
  ABJ(M) Theory},'' \href{http://dx.doi.org/10.1007/JHEP03(2012)062}{{\em JHEP}
  {\bf 1203} (2012)  062},
\href{http://arxiv.org/abs/1112.6299}{{\tt arXiv:1112.6299 [hep-th]}}.
P.~Caputa and B.~A.~E. Mohammed, ``{From Schurs to Giants in ABJ(M)},''
\href{http://arxiv.org/abs/1210.7705}{{\tt arXiv:1210.7705 [hep-th]}}.



\bibitem{Bianchi:2011rn}
M.~S. Bianchi, M.~Leoni, A.~Mauri, S.~Penati, C.~Ratti, {\em et al.}, ``{From
  Correlators to Wilson Loops in Chern-Simons Matter Theories},''
  \href{http://dx.doi.org/10.1007/JHEP06(2011)118}{{\em JHEP} {\bf 1106} (2011)
   118},
\href{http://arxiv.org/abs/1103.3675}{{\tt arXiv:1103.3675 [hep-th]}}.


\bibitem{Alday:2010zy}
L.~F. Alday, B.~Eden, G.~P. Korchemsky, J.~Maldacena, and E.~Sokatchev, ``{From
  correlation functions to Wilson loops},''
  \href{http://dx.doi.org/10.1007/JHEP09(2011)123}{{\em JHEP} {\bf 1109} (2011)
   123},
\href{http://arxiv.org/abs/1007.3243}{{\tt arXiv:1007.3243 [hep-th]}}.
B.~Eden, G.~P. Korchemsky, and E.~Sokatchev, ``{More on the duality
  correlators/amplitudes},''
  \href{http://dx.doi.org/10.1016/j.physletb.2012.02.014}{{\em Phys.Lett.} {\bf
  B709} (2012)  247--253},
\href{http://arxiv.org/abs/1009.2488}{{\tt arXiv:1009.2488 [hep-th]}}.
B.~Eden, P.~Heslop, G.~P. Korchemsky, and E.~Sokatchev, ``{The
  super-correlator/super-amplitude duality: Part I},''
\href{http://arxiv.org/abs/1103.3714}{{\tt arXiv:1103.3714 [hep-th]}}.
T.~Adamo, M.~Bullimore, L.~Mason, and D.~Skinner, ``{A Proof of the
  Supersymmetric Correlation Function / Wilson Loop Correspondence},''
  \href{http://dx.doi.org/10.1007/JHEP08(2011)076}{{\em JHEP} {\bf 1108} (2011)
   076},
\href{http://arxiv.org/abs/1103.4119}{{\tt arXiv:1103.4119 [hep-th]}}.
B.~Eden, P.~Heslop, G.~P. Korchemsky, and E.~Sokatchev, ``{The
  super-correlator/super-amplitude duality: Part II},''
\href{http://arxiv.org/abs/1103.4353}{{\tt arXiv:1103.4353 [hep-th]}}.

\bibitem{Engelund:2011fg}
O.~T. Engelund and R.~Roiban, ``{On correlation functions of Wilson loops,
  local and non-local operators},''
  \href{http://dx.doi.org/10.1007/JHEP05(2012)158}{{\em JHEP} {\bf 1205} (2012)
   158},
\href{http://arxiv.org/abs/1110.0758}{{\tt arXiv:1110.0758 [hep-th]}}.

\bibitem{Minahan:2008hf}
J.~A. Minahan and K.~Zarembo, ``{The Bethe ansatz for superconformal
  Chern-Simons},'' \href{http://dx.doi.org/10.1088/1126-6708/2008/09/040}{{\em
  JHEP} {\bf 09} (2008)  040},
\href{http://arxiv.org/abs/0806.3951}{{\tt arXiv:0806.3951 [hep-th]}}.

\bibitem{Escobedo:2010xs}
J.~Escobedo, N.~Gromov, A.~Sever, and P.~Vieira, ``{Tailoring Three-Point
  Functions and Integrability},''
  \href{http://dx.doi.org/10.1007/JHEP09(2011)028}{{\em JHEP} {\bf 1109} (2011)
   028},
\href{http://arxiv.org/abs/1012.2475}{{\tt arXiv:1012.2475 [hep-th]}}.

\bibitem{Foda:2011rr}
O.~Foda, ``{N=4 SYM structure constants as determinants},''
  \href{http://dx.doi.org/10.1007/JHEP03(2012)096}{{\em JHEP} {\bf 1203} (2012)
   096},
\href{http://arxiv.org/abs/1111.4663}{{\tt arXiv:1111.4663 [math-ph]}};
  O.~Foda and M.~Wheeler,
  ``{Slavnov determinants, Yang-Mills structure constants, and discrete KP},''
\href{http://arxiv.org/abs/1203.5621}{{\tt arXiv:1203.5621 [hep-th]}}.

\bibitem{Escobedo:2011xw}
J.~Escobedo, N.~Gromov, A.~Sever, and P.~Vieira, ``{Tailoring Three-Point
  Functions and Integrability II. Weak/strong coupling match},'' {\em JHEP} {\bf 1109}, 029 (2011),
  \href{http://arxiv.org/abs/1104.5501}{{\tt arXiv:1104.5501 [hep-th]}}. 

\bibitem{Bissi:2011ha}
A.~Bissi, T.~Harmark, and M.~Orselli, ``{Holographic 3-Point Function at One
  Loop},'' \href{http://dx.doi.org/10.1007/JHEP02(2012)133}{{\em JHEP} {\bf
  1202} (2012)  133},
\href{http://arxiv.org/abs/1112.5075}{{\tt arXiv:1112.5075 [hep-th]}}.


\bibitem{Zarembo:2010rr}
K.~Zarembo, ``{Holographic three-point functions of semiclassical states},''
  \href{http://dx.doi.org/10.1007/JHEP09(2010)030}{{\em JHEP} {\bf 1009} (2010)
   030}, \href{http://arxiv.org/abs/1008.1059}{{\tt arXiv:1008.1059 [hep-th]}}.
R.~A. Janik, P.~Surowka, and A.~Wereszczynski, ``{On correlation functions of
  operators dual to classical spinning string states},''
  \href{http://dx.doi.org/10.1007/JHEP05(2010)030}{{\em JHEP} {\bf 1005} (2010)
   030},
\href{http://arxiv.org/abs/1002.4613}{{\tt arXiv:1002.4613 [hep-th]}}.
E.~Buchbinder and A.~Tseytlin, ``{On semiclassical approximation for
  correlators of closed string vertex operators in AdS/CFT},''
  \href{http://dx.doi.org/10.1007/JHEP08(2010)057}{{\em JHEP} {\bf 1008} (2010)
   057},
\href{http://arxiv.org/abs/1005.4516}{{\tt arXiv:1005.4516 [hep-th]}}.
M.~S. Costa, R.~Monteiro, J.~E. Santos, and D.~Zoakos, ``{On three-point
  correlation functions in the gauge/gravity duality},''
  \href{http://dx.doi.org/10.1007/JHEP11(2010)141}{{\em JHEP} {\bf 1011} (2010)
   141}, \href{http://arxiv.org/abs/1008.1070}{{\tt arXiv:1008.1070 [hep-th]}}.



\bibitem{Aharony:2008ug}
O.~Aharony, O.~Bergman, D.~L. Jafferis, and J.~Maldacena, ``{$\CN=6$}
  superconformal {Chern-Simons-matter} theories, {M2-branes} and their gravity
  duals,'' {{\em JHEP} {\bf 0810}, 091 (2008)},
\href{http://arxiv.org/abs/0806.1218}{{\tt arXiv:0806.1218 [hep-th]}}.

\bibitem{Gromov:2008qe}
N.~Gromov and P.~Vieira, ``{The all loop AdS4/CFT3 Bethe ansatz},''
  \href{http://dx.doi.org/10.1088/1126-6708/2009/01/016}{{\em JHEP} {\bf 01}
  (2009)  016},
\href{http://arxiv.org/abs/0807.0777}{{\tt arXiv:0807.0777 [hep-th]}}.

\bibitem{Arutyunov:2008if}
  G.~Arutyunov and S.~Frolov,
  ``Superstrings on AdS(4) x CP**3 as a Coset Sigma-model,''
  \href{http://dx.doi.org/10.1088/1126-6708/2008/09/129}{{\em JHEP} {\bf 11},
(2008)  089},
\href{http://arxiv.org/abs/0806.4940}{{\tt arXiv:0806.4940 [hep-th]}};
  B.~Stefanski, jr,
  ``Green-Schwarz action for Type IIA strings on AdS(4) x CP**3,''
  Nucl.\ Phys.\ B {\bf 808} (2009) 80
\href{http://arxiv.org/abs/0806.4948}{{\tt arXiv:0806.4948 [hep-th]}}.

\bibitem{Minahan:2002ve}
J.~A. Minahan and K.~Zarembo, ``The {Bethe-ansatz} for {$\CN = 4$} super
  {Yang-Mills},'' {\em JHEP} {\bf 03} (2003)  013,
\href{http://arxiv.org/abs/hep-th/0212208}{{\tt hep-th/0212208}}.
N.~Beisert, C.~Kristjansen, and M.~Staudacher, ``The dilatation operator of
  {$\CN = 4$} super {Yang-Mills} theory,'' {\em Nucl. Phys.} {\bf B664} (2003)
  131--184,
\href{http://arxiv.org/abs/hep-th/0303060}{{\tt hep-th/0303060}}.
N.~Beisert and M.~Staudacher, ``Long-range {${PSU}(2,2|4)$} {Bethe} ansaetze
  for gauge theory and strings,'' {\em Nucl. Phys.} {\bf B727} (2005)  1--62,
\href{http://arxiv.org/abs/hep-th/0504190}{{\tt hep-th/0504190}}.
N.~Beisert, B.~Eden, and M.~Staudacher, ``{Transcendentality and crossing},''
  {\em J. Stat. Mech.} {\bf 0701} (2007)  P021,
\href{http://arxiv.org/abs/hep-th/0610251}{{\tt hep-th/0610251}}.


\bibitem{Mandal:2002fs}
G.~Mandal, N.~V. Suryanarayana, and S.~R. Wadia, ``{Aspects of semiclassical
  strings in AdS(5)},''
  \href{http://dx.doi.org/10.1016/S0370-2693(02)02424-3}{{\em Phys.Lett.} {\bf
  B543} (2002)  81--88},
\href{http://arxiv.org/abs/hep-th/0206103}{{\tt arXiv:hep-th/0206103
  [hep-th]}}.
I.~Bena, J.~Polchinski, and R.~Roiban, ``{Hidden symmetries of the {$\ads_5
  \times S^5$}superstring},'' {\em Phys. Rev.} {\bf D69} (2004)  046002,
\href{http://arxiv.org/abs/hep-th/0305116}{{\tt hep-th/0305116}}.


\bibitem{Pereira:2012}
R.\ Pereira, ``Structure constants in N=4 SYM and ABJM theory,''
Master Thesis, The Perimeter Institute, 2012, unpublished. 

\bibitem{Grignani:2008is}
G.~Grignani, T.~Harmark, and M.~Orselli, ``{The SU(2) x SU(2) sector in the
  string dual of N=6 superconformal Chern-Simons theory},''
  \href{http://dx.doi.org/10.1016/j.nuclphysb.2008.10.019}{{\em Nucl. Phys.}
  {\bf B810} (2009)  115--134},
\href{http://arxiv.org/abs/0806.4959}{{\tt arXiv:0806.4959 [hep-th]}}.
D.~Astolfi, V.~G.~M. Puletti, G.~Grignani, T.~Harmark, and M.~Orselli,
  ``{Finite-size corrections in the SU(2) $\times$ SU(2) sector of type IIA
  string theory on {$\ads_4 \times \C P^3$}},''
  \href{http://dx.doi.org/10.1016/j.nuclphysb.2008.10.020}{{\em Nucl. Phys.}
  {\bf B810} (2009)  150--173},
\href{http://arxiv.org/abs/0807.1527}{{\tt arXiv:0807.1527 [hep-th]}}.


\bibitem{Kostov:2012jr}
I.~Kostov, ``{Classical Limit of the Three-Point Function of N=4 Supersymmetric
  Yang-Mills Theory from Integrability},''
  \href{http://dx.doi.org/10.1103/PhysRevLett.108.261604}{{\em Phys.Rev.Lett.}
  {\bf 108} (2012)  261604},
\href{http://arxiv.org/abs/1203.6180}{{\tt arXiv:1203.6180 [hep-th]}}.
I.~Kostov, ``{Three-point function of semiclassical states at weak coupling},''
\href{http://arxiv.org/abs/1205.4412}{{\tt arXiv:1205.4412 [hep-th]}}.


\bibitem{Georgiou:2011qk}
G.~Georgiou, ``{SL(2) sector: weak/strong coupling agreement of three-point
  correlators},'' \href{http://dx.doi.org/10.1007/JHEP09(2011)132}{{\em JHEP}
  {\bf 1109} (2011)  132},
\href{http://arxiv.org/abs/1107.1850}{{\tt arXiv:1107.1850 [hep-th]}}.

\bibitem{Bastianelli:1999bm}
F.~Bastianelli and R.~Zucchini, ``{Bosonic quadratic actions for 11-D
  supergravity on AdS(7/4) x S(4/7)},''
  \href{http://dx.doi.org/10.1088/0264-9381/16/11/313}{{\em Class.Quant.Grav.}
  {\bf 16} (1999)  3673--3684},
\href{http://arxiv.org/abs/hep-th/9903161}{{\tt arXiv:hep-th/9903161
  [hep-th]}}.

\bibitem{Frolov:2003xy}
S.~Frolov and A.~A. Tseytlin, ``{Rotating string solutions: {AdS/CFT} duality
  in non- supersymmetric sectors},'' {\em Phys. Lett.} {\bf B570} (2003)
  96--104,
\href{http://arxiv.org/abs/hep-th/0306143}{{\tt hep-th/0306143}}.

\bibitem{Kruczenski:2003gt}
M.~Kruczenski, ``{Spin chains and string theory},'' {\em Phys. Rev. Lett.} {\bf
  93} (2004)  161602,
\href{http://arxiv.org/abs/hep-th/0311203}{{\tt hep-th/0311203}}.

\bibitem{Harmark:2008gm}
T.~Harmark, K.~R. Kristjansson, and M.~Orselli, ``{Matching gauge theory and
  string theory in a decoupling limit of AdS/CFT},'' {\em JHEP} {\bf 0902}, 027 (2009),
\href{http://arxiv.org/abs/0806.3370}{{\tt arXiv:0806.3370 [hep-th]}}.

\bibitem{Bissi:2012vx}
A.~Bissi, G.~Grignani, and A.~Zayakin, ``{The SO(6) Scalar Product and
  Three-Point Functions from Integrability},''
\href{http://arxiv.org/abs/1208.0100}{{\tt arXiv:1208.0100 [hep-th]}}.

\bibitem{Ahn:2012uv}
C.~Ahn, O.~Foda, and R.~I. Nepomechie, ``{OPE in planar QCD from
  integrability},'' \href{http://dx.doi.org/10.1007/JHEP06(2012)168}{{\em JHEP}
  {\bf 1206} (2012)  168},
\href{http://arxiv.org/abs/1202.6553}{{\tt arXiv:1202.6553 [hep-th]}}.

\bibitem{Gromov:2012vu}
N.~Gromov and P.~Vieira, ``{Quantum Integrability for Three-Point Functions},''
\href{http://arxiv.org/abs/1202.4103}{{\tt arXiv:1202.4103 [hep-th]}}.


\bibitem{Grignani:2012yu}
G.~Grignani and A.~Zayakin, ``{Matching Three-point Functions of BMN Operators
  at Weak and Strong coupling},''
  \href{http://dx.doi.org/10.1007/JHEP06(2012)142}{{\em JHEP} {\bf 1206} (2012)
   142},
\href{http://arxiv.org/abs/1204.3096}{{\tt arXiv:1204.3096 [hep-th]}}.
G.~Grignani and A.~Zayakin, ``{Three-point functions of BMN operators at weak
  and strong coupling II. One loop matching},''
  \href{http://dx.doi.org/10.1007/JHEP09(2012)087}{{\em JHEP} {\bf 1209} (2012)
   087},
\href{http://arxiv.org/abs/1205.5279}{{\tt arXiv:1205.5279 [hep-th]}}.


\bibitem{Gromov:2012uv}
N.~Gromov, P.~Vieira, and P.~Vieira, ``{Tailoring Three-Point Functions and
  Integrability IV. Theta-morphism},''
\href{http://arxiv.org/abs/1205.5288}{{\tt arXiv:1205.5288 [hep-th]}}.

\bibitem{Serban:2012dr}
D.~Serban, ``{A note on the eigenvectors of long-range spin chains and their
  scalar products},''
\href{http://arxiv.org/abs/1203.5842}{{\tt arXiv:1203.5842 [hep-th]}}.

\bibitem{Kruczenski:2004kw}
M.~Kruczenski, A.~V. Ryzhov, and A.~A. Tseytlin, ``{Large spin limit of
  {$\ads_5\times S^5$} string theory and low energy expansion of ferromagnetic
  spin chains},'' {\em Nucl. Phys.} {\bf B692} (2004)  3--49,
\href{http://arxiv.org/abs/hep-th/0403120}{{\tt hep-th/0403120}}.

\end{thebibliography}
\end{document}